\documentclass[prx,amsmath,amssymb,twocolumn,longbibliography]{revtex4-1}

\usepackage{graphicx}
\usepackage{dcolumn}
\usepackage{bm}
\usepackage[utf8]{inputenc}
\usepackage[english]{babel}
\usepackage{amsfonts}
\usepackage{hyperref}
\usepackage{physics}
\usepackage{epstopdf}
\usepackage{mathtools}
\usepackage[capitalise]{cleveref}
\usepackage{float}
\usepackage[space]{grffile}
\usepackage{color}
\usepackage[normalem]{ulem}
\usepackage{subfigure}
\usepackage{natbib}
\usepackage{amsmath}
\usepackage{gensymb}

\def\avg #1{\left< #1 \right>}

\def\diamondcomma{\ \raise.3ex\hbox{$\diamond$}\kern-0.4em\lower.7ex\hbox{$,$}\ }
\def\lesssim{\ \raise.3ex\hbox{$<$}\kern-0.8em\lower.7ex\hbox{$\sim$}\ }
\def\gesim{\ \raise.3ex\hbox{$>$}\kern-0.8em\lower.7ex\hbox{$\sim$}\ }

\begin{document}

    \title{Non-reciprocal frustration: time crystalline order-by-disorder phenomenon and a spin-glass-like state
    }
    \author{Ryo Hanai}
    \email{ryo.hanai@yukawa.kyoto-u.ac.jp} 
	\affiliation{Center for Gravitational Physics and Quantum Information, Yukawa Institute for Theoretical Physics, Kyoto University, Kyoto 606-8502, Japan}	\affiliation{Asia Pacific Center for Theoretical Physics, Pohang 37673, Korea}
    
	\date{\today}


    \begin{abstract}
    Active systems are comprised of constituents with interactions that are generically non-reciprocal in nature. Such non-reciprocity often gives rise to situations where conflicting objectives exist, such as in the case of a predator pursuing its prey, while the prey attempts to evade capture. 
    This situation is somewhat reminiscent of those encountered in geometrically frustrated systems where conflicting objectives also exist, which result in the absence of configurations that simultaneously minimize all interaction energies. 
    In the latter, a rich variety of exotic phenomena 
    are known to arise due to the presence of accidental degeneracy of ground states. 
    In this paper, we establish a \textit{direct} analogy between 
    these two classes of systems.
    The analogy is based on the observation that non-reciprocally interacting systems with anti-symmetric coupling and geometrically frustrated systems have in common that they both exhibit marginal orbits, which can be regarded as a dynamical system counterpart of accidentally degenerate ground states. 
    The former is shown by proving a Liouville-type theorem.
    These ``accidental degeneracies'' of orbits are shown to often get ``lifted'' by stochastic
    noise or weak random disorder 
    due to the emergent ``entropic force''
    to give rise to a noise-\textit{induced} spontaneous symmetry breaking, in a similar manner to the order-by-disorder phenomena known to occur in geometrically frustrated systems.
    Furthermore, we report numerical evidence of a non-reciprocity-induced spin-glass-like state that exhibits a short-ranged spatial correlation (with stretched exponential decay) and an algebraic temporal correlation
    associated with the aging effect.
    Our work establishes an unexpected connection between the physics of complex magnetic materials and non-reciprocal matter, offering a fresh and valuable perspective for comprehending the latter.
	\end{abstract}

\maketitle

\section{Introduction}

	\begin{figure*}[t]
    \centering
    \includegraphics[width=0.7\linewidth,keepaspectratio]{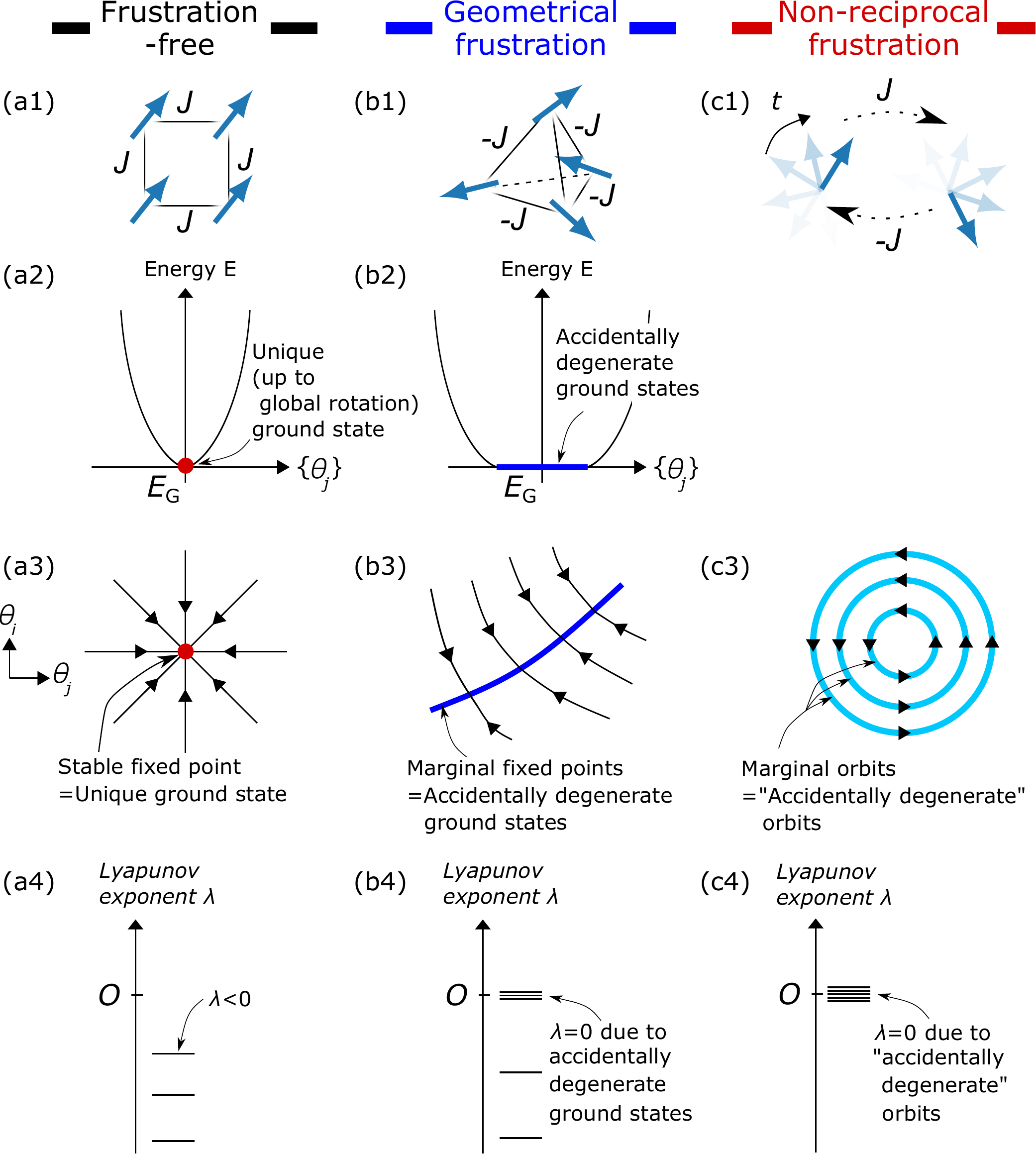}
	\caption{
    \textbf{Geometrical and non-reciprocal frustration and the emergence of ``accidental degeneracy'' of orbits.}
    (a1)-(c1) Examples of systems with (a1) no frustration, (b1) geometrical frustration, and (c1) anti-symmetric non-reciprocal interactions. 
    (a1),(b2) Schematic description of energy $E$ as a function of the spin configuration $\{\theta_j\}$. Here, $E_{\rm G}$ is the ground state energy and we have omitted the degeneracy that trivially arises from the global rotation symmetry, for clarity of the figure. 
    Note that the energy $E$ is not defined for the non-reciprocal case.
    (a3)-(c3) Orbits.
    (a4)-(c4) Lyapunov exponents $\lambda$. 
	(a) In frustration-free systems, since fixing the angle of one spin would determine all other spin configurations to minimize the energy of the system, the ground state is unique up to global symmetry. 
	As a result, the system converges into a unique stable fixed point (red point in (a3))), which gives negative Lyapunov exponents $\lambda<0$. 
	(b) Geometrically frustrated systems, on the other hand, often exhibit accidentally degenerate ground states because of the underconstrained degrees of freedom.
	The presence of the degenerate ground states implies that there is a direction in which a restoring force (torque) is absent.
	This means that the accidentally degenerate ground states corresponds to marginal fixed points in the language of dynamical systems (blue line in (b3)) that have zero Lyapunov exponent(s) $\lambda=0$.
	(c) In non-reciprocally frustrated systems with perfect non-reciprocity $J_{ij}=-J_{ji}$, the spins start a chase-and-runaway motion that corresponds to marginal orbits with zero Lyapunov exponents $\lambda=0$, arising due to the Liouville-type theorem (Eq.~\eqref{eq: Liouville's theorem}).  These orbits can be regarded as the dynamical counterpart of the ground state accidental degeneracy of geometrically frustrated systems that also have zero Lyapunov exponents $\lambda=0$.
	}
	\label{fig: frustration}
	\end{figure*}
There is a current surge of interest in the physics of non-reciprocally interacting active systems \cite{Bowick2022, Shankar2022}.
Non-reciprocal interaction refers to an asymmetry in the interaction between two or more entities in which the action and reaction are not equal. 
This phenomenon arises generically whenever the system is coupled to a non-equilibrium environment, and is, therefore, ubiquitous in nature \cite{Ivlev2015, Uchida2010, Soto2014, Saha2019, Yifat2018, Parker2020, Zhang2021, Tan2022, Fruchart2021, Neubert1997, Kerr2002, Reichenbach2007, Rieger_eco1989, Allesina2012, Bunin2017, Hong2011, Wilson1972, Sompolinsky1986, Rieger1988, Rieger1989, Brandenbourger2019, Ashida2021, Metelmann2015, Hatano1996}.
The importance of non-reciprocal interaction has been extensively acknowledged in various scientific disciplines, ranging from active matter~\cite{Ivlev2015, Uchida2010, Soto2014, Saha2019, Yifat2018, Parker2020, Tan2022, Zhang2021, Fruchart2021}, ecology~\cite{Neubert1997, Kerr2002, Reichenbach2007, Rieger_eco1989, Allesina2012, Bunin2017}, social science~\cite{Hong2011}, neuroscience~\cite{Wilson1972, Sompolinsky1986, Rieger1988, Rieger1989}, robotics~\cite{Brandenbourger2019}, to open quantum systems ~\cite{Ashida2021, Metelmann2015, Hatano1996}. 
More recently, researchers have found that non-reciprocal interaction significantly impacts the collective behavior of many-body systems \cite{Hatano1996, Fruchart2021, You2020, Saha2020, Weis2022, Zhang2021, Pakard2022, Dadhichi2020, Loos2022, Scheibner2020, Tan2022}.
The effects include the emergence of odd elasticity \cite{Scheibner2020,Tan2022}, non-reciprocal phase transitions \cite{Fruchart2021,You2020,Saha2020}, and long-ranged order in two spatial dimensions \cite{Dadhichi2020,Loos2022}.

In the presence of non-reciprocal interactions, conflicting objectives often arise. As a simple example, consider a case where agent A attracts agent B while B repulses A. In such a situation, no configurations can satisfy both agents, as A seeks to be close to B while B desires the opposite. 
This situation is, to some extent, analogous to situations encountered in geometrically frustrated systems.
Geometrically frustrated systems are defined as systems that cannot satisfy the constituents' ``desire'' to minimize all interaction energy \cite{Toulouse1977,Moessner2006}.
(See Figs.~\ref{fig: frustration}(a1) and (b2) for a typical example of a frustration-free and geometrically frustrated system, respectively.)
In other words, these systems do not have any configurations that can make all constituents `happy', similar to the non-reciprocally interacting systems.
This means that at least some constituents must compromise for global optimization. 
As there can be many ways to achieve this, geometrical frustrated systems often exhibit accidentally degenerate ground states
(Figs.~\ref{fig: frustration}(a2),(b2)).
This not only makes the system extremely sensitive to external perturbations but also gives rise to various exotic phenomena, such as order-by-disorder phenomena (OBDP) \cite{Villain1980,Henley1989,Moessner1998a,Moessner1998b,Bergman2007}, spin glass \cite{Edwards1975,Sherrington1975,Parisi1979,Sompolinsky1982,Cugliandolo1993,Fisher1998a,Fisher1998b}, spin ice \cite{Castelnovo2012}, and quantum \cite{Savary2016} and classical \cite{Moessner1998a,Moessner1998b} spin liquids, in and out of equilibrium \cite{Wan2017, Bittner2020, Jin2022, Yue2023}.

This raises the question of whether non-reciprocally interacting systems can also give rise to phenomena similar to those induced by geometrical frustration.
Phrased differently, is there a counterpart of accidentally degenerate ground states in non-reciprocal systems, that were the origin of these exotic phenomena?
At first glance, it seems highly unlikely, due to a crucial difference from geometrically frustrated systems: the absence of the notion of energy in non-reciprocal systems \cite{Fruchart2021}.
Consequently, accidentally degenerate ground states cannot be defined.
Moreover, non-reciprocally interacting agents typically start a ``chase-and-runaway'' motion that cannot be described in terms of energy minimization 
(Fig.~\ref{fig: frustration}(c1)), unlike in geometrically frustrated systems where they settle in a compromised configuration.
As such, the two types of frustration appear to have no further connection in their phenomenology beyond the vague resemblance mentioned earlier.

Despite these fundamental differences, in this paper, we establish a \textit{direct} analogy between the two types of frustration. 
This is achieved by pointing out a crucial common feature shared between 
geometrically frustrated systems and non-reciprocally interacting systems with anti-symmetric coupling: the presence of marginal orbits characterized by zero Lyapunov exponents that do not originate from symmetry (see Figs. \ref{fig: frustration}(a3)-(c3) and (a4)-(c4)). 
In the case of geometric frustration (Fig.~\ref{fig: frustration}(b1)), the accidentally degenerate ground state (Fig.~\ref{fig: frustration}(b2)) corresponds to marginal \textit{fixed points} (Fig.~\ref{fig: frustration}(b3)) in the dynamical system language.
The presence of marginal orbits in non-reciprocally interacting cases, on the other hand, is supported by a Liouville-type theorem that holds in the anti-symmetric coupling limit.
In contrast to the marginal fixed points in the geometrically frustrated systems, the marginal orbits in the non-reciprocal systems are generally time-dependent (Fig.~\ref{fig: frustration}(c3)), reflecting their non-equilibrium nature. 
The emerging marginal orbits in the latter can therefore be viewed as the dynamical counterparts of the accidentally degenerate ground states.

We show that these ``accidentally degenerate'' orbits often get ``lifted'' by stochastic noise or quenched disorder.
This leads to the emergence of OBDP, where noise or quenched disorder \textit{induce} order instead of destroying them, which is opposite from what one usually expects. 
This effect is attributed to the emergence of the ``entropic (disorder-induced) force'' that naturally arises in the presence of accidental degeneracy combined with stochasticity.
We show that this ``entropic (disorder-induced) force'' can trigger noise(disorder)-\textit{induced} spontaneous symmetry breaking via non-reciprocal phase transition \cite{Fruchart2021}.
In addition, we provide numerical evidence that a spin glass-like state emerges in a randomly coupled spin chain with non-reciprocal interaction but has no geometric frustration. 
There, we observe a power-law decay of a time-correlation function with a clear sign of aging, while the spatial correlation function is found to be short-ranged (stretched exponential decay).
These findings establish an unexpected connection between the seemingly unrelated fields of complex magnetic materials and non-reciprocal matter.
Our results may have applications in the field of active matter and biological systems and offer a novel design principle for the robotic metamaterial.

The paper is organized as follows. In Sec. \ref{sec: accidental degeneracy}, we 
draw a direct analogy between geometrically frustrated systems and non-reciprocally interacting systems with anti-symmetric coupling
by proving that marginal orbits, which can be regarded as a dynamical counterpart of accidental degeneracy, generically arise in both classes of systems.
This is shown by proving a Liouville-type theorem that holds in this limit. 
In Sec. \ref{sec:order-by-disorder}, we demonstrate that this ''accidental degeneracy'' of orbits typically gets ``lifted'' by stochastic noise or quenched disorder, giving rise to time-crystalline OBDP.
In Sec. \ref{sec:spin glass}, we show numerically that a state analogous to a spin glass state emerges when non-reciprocity is introduced in their coupling in a one-dimensional randomly coupled spin chain.
In Sec. \ref{sec:conclusion}, we summarize our paper and discuss the outlook.

\section{The emergence of  ``accidental degeneracy'' of orbits}
\label{sec: accidental degeneracy}

In this paper, 
although the concepts we introduce should be valid for a more general class of models (See Appendix \ref{Appendix: Liouville-type theorem}), 
for concreteness, we consider 
dissipatively coupled classical XY spin  
systems with their spin angle $\theta=(\theta_1,...,\theta_N)$ dynamics governed by 
\begin{eqnarray}
    \label{eq: XY-model}
    \dot\theta_i = -\sum_{j=1}^N J_{ij}\sin(\theta_i - \theta_j),
\end{eqnarray}
which generically has a non-reciprocal coupling $J_{ij}\ne J_{ji}$.
The effect of stochastic noise will be addressed later.
This dynamical system (Eq.~\eqref{eq: XY-model})  is invariant under global rotation $\theta_i\rightarrow \theta_i + \chi$ (where $\chi$ is a real constant).

Let us first briefly review the reciprocal coupling case $J_{ij}=J_{ji}$ with and without geometrical frustration. 
In such systems, Eq.~\eqref{eq: XY-model} can be rewritten using a derivative of a energy $E(\theta)$ as $\dot\theta_i = -
     \partial E(\theta)/\partial\theta_i$,
where
\begin{eqnarray}
\label{eq: energy}
    E(\theta) = 
    -\sum_{i,j} J_{ij}\cos(\theta_i-\theta_j). 
\end{eqnarray}
As a result, the system is driven towards the (local) minimum of the energy $E$.

Frustration-free systems are systems that have ground states that minimize the energy $E$ of Eq.~\eqref{eq: energy} \textit{term by term}
\cite{Toulouse1977} (Fig. \ref{fig: frustration}(a)).
In this case, the ground state configuration is uniquely determined by fixing one of the spin angle. 
Therefore, these systems only have a ground state degeneracy that is trivially due to the rotation symmetry of the dynamical system 
(Fig.~\ref{fig: frustration}(a2)).
The dynamical system \eqref{eq: XY-model} therefore has a unique stable fixed point (Fig.~\ref{fig: frustration}(a3)) when  
regarding the orbits identical up to global rotation as the same orbit.
In this case, all Lyapunov exponents would be negative $\lambda_i<0$ (Fig.~\ref{fig: frustration}(a4)) except for the zero modes arising from the global rotation symmetry (i.e. the Nambu-Goldstone mode, which we omitted in Fig.~\ref{fig: frustration}(a4)).

In contrast, in geometrically frustrated systems (Fig.~\ref{fig: frustration}(b)),  
there are no configurations that simultaneously minimize all the interaction terms \cite{Toulouse1977}.
In such a situation, the ground state configuration is often underconstrained \cite{Moessner1998a,Moessner1998b,Moessner2006}, causing the emergence of an accidental degeneracy of ground states (Fig.~\ref{fig: frustration}(b2)) that does not stem from their underlying symmetry. 
Which ground state the system ultimately converges to depends on its initial condition.
No restoring force would be applied in the direction in-plane to the accidentally degenerate ground state manifold.
In the language of dynamical systems, this implies the existence of marginal fixed points (Fig.~\ref{fig: frustration}(b3)) that indicate the presence of zero Lyapunov exponent(s) $\lambda=0$ (Fig.~\ref{fig: frustration}(b4)) (in addition to the zero Lyapunov exponent trivially arising from the Nambu-Goldstone mode).





We show below that the non-reciprocally interacting system with anti-symmetric coupling $J_{ij}=-J_{ji}$ (which we refer to below as `\textit{perfectly non-reciprocal}') has exactly the same feature: the existence of marginal orbits with zero Lyapunov exponents $\lambda=0$ (Fig. \ref{fig: frustration}(c)).
In this situation, the distribution function $\rho(\theta)$ is found to stay constant along any trajectory (See Appendix \ref{Appendix: Liouville-type theorem} for the proof.), i.e.,
\begin{eqnarray}
\label{eq: Liouville's theorem}
    \frac{d\rho}{dt}
    = \frac{\partial\rho}{\partial t}
    +\sum_{i}
    \frac{\partial \rho}{\partial \theta_i}
    \dot \theta_i
    =0,
\end{eqnarray}
in a similar manner to Liouville's theorem of Hamiltonian systems.
Note that a similar theorem holds for non-reciprocally interacting Heisenberg models, oscillators with phase-delayed interactions \cite{Sakaguchi1986} (that well-describes biased Josephson junctions arrays \cite{Wiesenfeld1996,  Wiesenfeld1998} and microscopic rotors \cite{Uchida2010}), and non-reciprocally interacting particles (that describes e.g., complex plasma \cite{Ivlev2015} and chemically \cite{Soto2014, Saha2019} and optically active colloidal matter \cite{Yifat2018, Parker2020}), as shown in Appendix \ref{Appendix: Liouville-type theorem} 
(see also Ref.~\cite{Hofbauer1994} for a similar relation known in the context of evolutionary game theories).
The conservation of phase volume $dV=\rho\prod_i d\theta_i$ of Eq.~\eqref{eq: Liouville's theorem} means that the dynamics are dissipationless and 
the sum of all Lyapunov exponents is zero $\sum_{i=1}^N\lambda_i = 0$.
In the absence of chaos $\lambda_i\le 0$, this makes all Lyapunov exponents vanish $\lambda_i=0$ (Fig. \ref{fig: frustration}(c4)), which, generically, implies the emergence of marginal orbits described schematically in Fig. \ref{fig: frustration}(c3). 
Which orbit the system actually takes depends on the initial condition, in an identical situation to the geometrically frustrated case.

We interpret these marginal orbits as the emergence of ``accidental degeneracy" caused by non-reciprocal frustration. 
This degeneracy is accidental, in the sense that they do not originate from the global symmetry or topology of the dynamical system (Eq.~\eqref{eq: XY-model}), in direct analogy to those of geometrical frustration.
The difference lies both in its physical origin and the consequence: in the non-reciprocal (geometrical) frustration case, the degeneracy comes from Liouville's theorem (underconstrained degrees of freedom \cite{Moessner1998a, Moessner1998b, Moessner2006}) and the resulting marginal orbits are 
typically time-dependent (static).

\begin{figure}[t]
\centering
\includegraphics[width=0.55\linewidth,keepaspectratio]{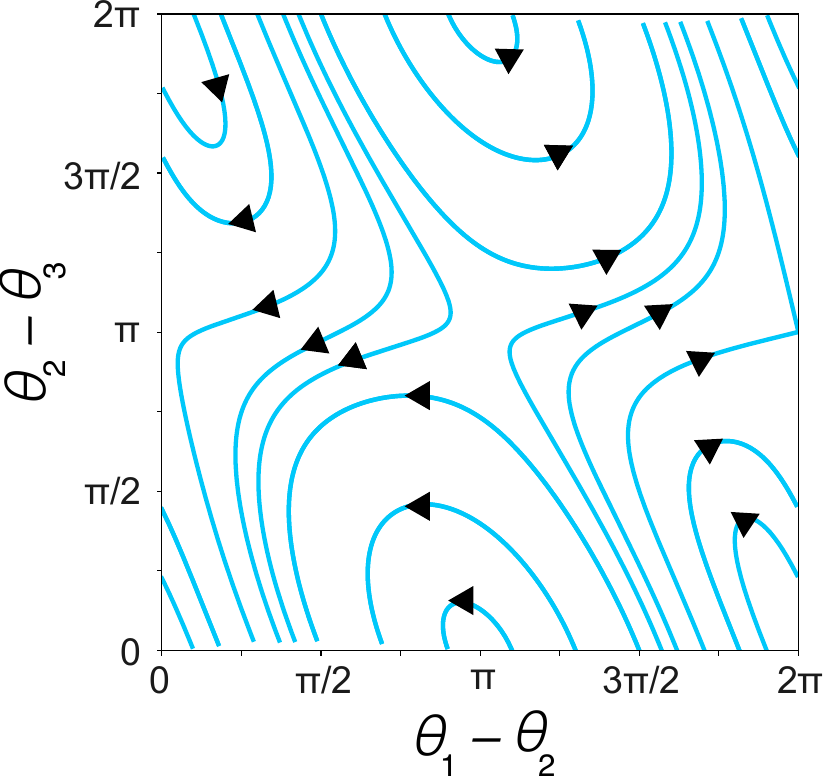}
\caption{
\textbf{Marginal orbits in perfectly non-reciprocal three spin system.}
We set $J_{12}=-J_{21}=3,J_{23}=-J_{32}=-1,J_{31}=-J_{13}=2$.
}
\label{fig: three spins}
\end{figure}

Take a two-spin perfectly non-reciprocal system $J_{12}=-J_{21}=J_-$ as the simplest example \cite{Fruchart2021,Baek2018}. One can readily find an analytical solution to the center-of-mass angle $\Theta=(\theta_1+\theta_2)/2$ and the difference $\Delta\theta=\theta_1 - \theta_2$ for a given initial condition $\theta_{i=1,2}(t=0)$ as   
\begin{eqnarray}
\label{eq: two components}
    \Theta(t) =  - J_- t \sin[\Delta\theta(0)] , 
    \qquad 
    \Delta\theta(t)=\Delta\theta(0). 
\end{eqnarray}
As expected, the system exhibits marginal periodic orbits, where the speed and direction of the drift of the center-of-mass angle $\Theta$ are determined by the initial condition of $\Delta\theta$ that stays constant. 
The numerical solution of a three-spin perfectly non-reciprocal system is depicted in Fig. \ref{fig: three spins} as another example, where we similarly find marginal periodic orbits.

Accidental degeneracy is usually associated with fine-tuning of parameters. Here, in non-reciprocally frustrated systems, the emergence of marginal orbits relies on the fine-tuning of the coupling to be perfectly non-reciprocal $J_{ij}=-J_{ji}$.
Once the coupling strength deviates from this limit, the marginal orbits would generically turn into (un)stable orbits, corresponding to the `lifting' of degeneracy.
This situation is in parallel to the geometrical frustration case where the degeneracy is contingent on the coupling strength being identical $J_{ij}=J$ \cite{Moessner2006}. 


So far, we considered cases  where  \textit{all} spins are perfectly non-reciprocally interacting $J_{ij}=-J_{ji}$, where we have shown that  Liouville-type theorem Eq.~\eqref{eq: Liouville's theorem} holds in such cases.
This means that there is absolutely no dissipation occurring in the system:
\textit{any} initial state will exhibit a marginal orbit that conserves the phase volume in this case.
In some sense, this is similar to systems with a \textit{constant} energy $E(\theta)={\rm const.}$, where \textit{all} states are trivially in the ground state manifold. 
This stands in contrast to generic geometrically frustrated systems, where only a small subset of states reside in the ground state manifold.
In such systems, a typical initial state would \textit{relax} to a state that corresponds to a marginal fixed point, which is in contrast to systems with constant energy $E(\theta)={\rm const.}$ where no relaxation occur.

In what follows, we show that there is a class of non-reciprocal systems
where a generic initial state relaxes to an orbit that is marginal, making the analogy to geometrically frustrated systems even more direct. 
Namely, we consider a system that is separated into communities that interact non-reciprocally between different communities but ferromagnetically within the same community,
\begin{eqnarray}
\label{eq: many communities}
    \dot\theta_i^a 
    = \sum_b\sum_j J_{ij}^{ab}\sin(\theta_j^b-\theta_i^a).
\end{eqnarray}
Here, $a,b$ labels the community, and $i,j$ labels the spins in the community and the intra-community coupling is ferromagnetic $J_{ij}^{aa}>0$.
In such a situation, 
the spins in the intra-communities would eventually align $\theta_i^a = \phi_a$ to give
\begin{eqnarray}
\label{eq: phase order dynamics deterministic}
    \dot\phi_a = \sum_b j_{ab}\sin(\phi_b-\phi_a),
\end{eqnarray}
in the long time limit, which has an identical form to Eq.~\eqref{eq: XY-model}
when the inter-community coupling $j_{ab}=\sum_{j}J_{ij}^{ab}~(a\ne b)$ is $i$-independent. 
Therefore, following the same logic as before, the system would exhibit marginal orbits with zero Lyapunov exponents in the perfectly non-reciprocal \textit{inter-}community coupling
$j_{ab}=-j_{ba}$.
The difference from the systems considered before is the presence of dissipative processes towards this attractor due to the ferromagnetic \textit{intra-}community interactions $J_{ij}^{aa}>0$.
In the next section, we will show that such a relaxation process combined with the presence of marginal orbits indeed plays a crucial role in the emergence of a counterintuitive phenomenon called the OBDP.
\section{Time crystalline order-by-disorder phenomena}
\label{sec:order-by-disorder}

\begin{figure}[t]
\centering
\includegraphics[width=1\linewidth,keepaspectratio]{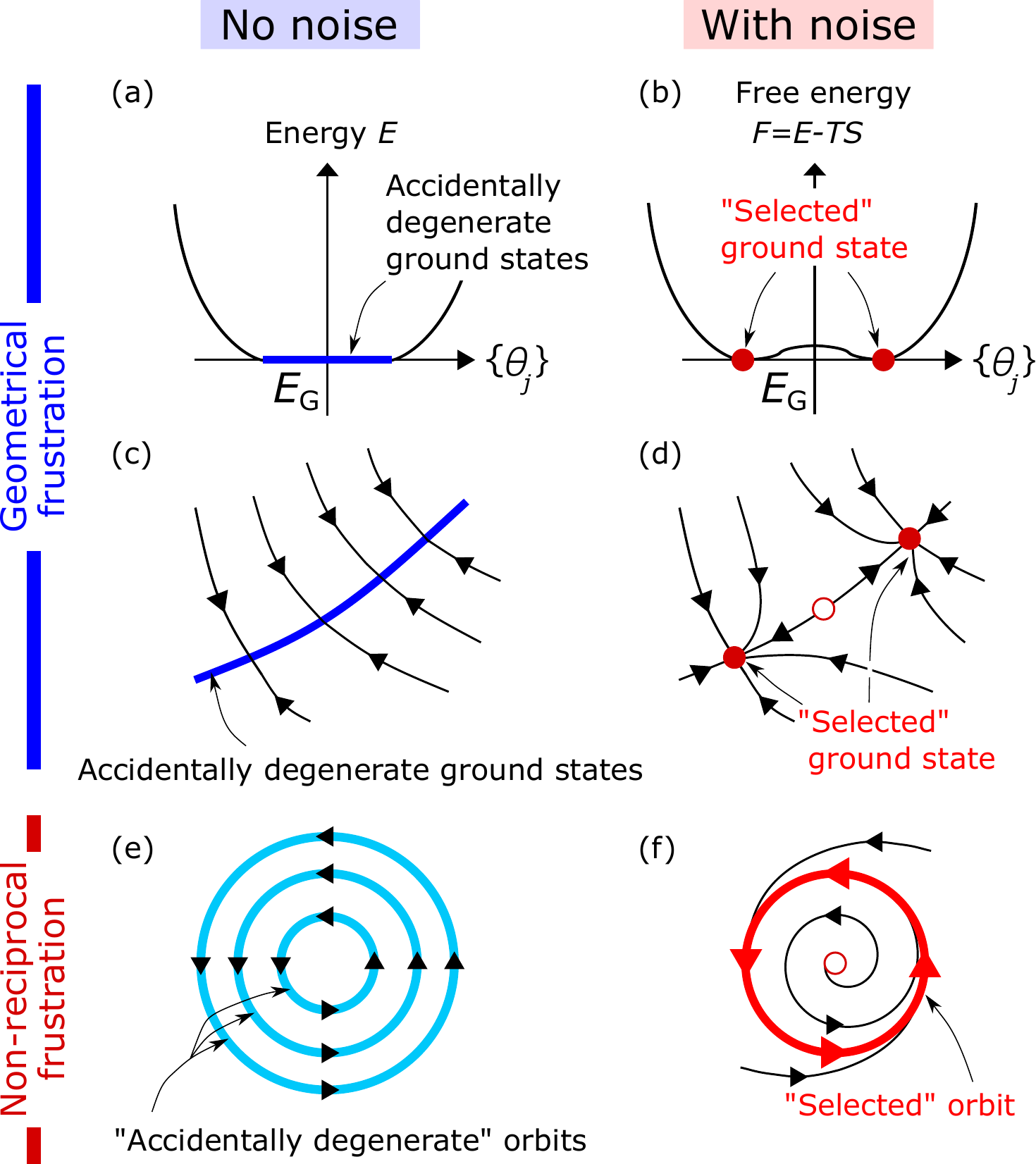}
\caption{
\textbf{The concept of order by disorder phenomena in equilibrium and their generalization to non-reciprocally interacting systems.}
(a),(b) Energy and free energy profiles in a geometrically frustrated system  (a) without noise (i.e., zero temperature $T = 0$) and (b) with noise (i.e., finite temperature $T > 0$) in a geometrically frustrated system. (c),(d) Orbits in a geometrically frustrated system without (c) and with (d) thermal noise. 
At $T = 0$, the system exhibits ground state degeneracy, while the presence of thermal noise at $T > 0$ lifts this degeneracy through entropic forces. (e),(f) Orbits in a non-reciprocally interacting system without (e) and with (f) noise.
Similarly to the geometrically frustrated system counterpart, ``entropic force'' selects one of the orbits to give rise to OBDP.
}
\label{fig: order by disorder schematic}
\end{figure}

Having established that non-reciprocal interaction gives an alternative route from geometrical frustration to generating ``accidental degeneracy'' of orbits (i.e., marginal orbits),
we now investigate their impact on the many-body properties of the system.
In geometrically frustrated systems, a paradigmatic example of a phenomenon emerging from such accidental degeneracy
is the OBDP \cite{Villain1980, Henley1989, Moessner1998a, Moessner1998b, Bergman2007}.
As the degeneracy generated by frustration  is not protected by symmetry nor topology, it is fragile, not only against external perturbations 
but also against disorders such as thermal noise or weak random potential.
As a result, the degeneracy often gets lifted and ends up, perhaps counter-intuitively, in a more ordered state than that of the clean system.
This is known as the OBDP.

In this section, we show that an analogous phenomenon arises in the non-reciprocally interacting many-body systems as well, with the peculiarity that the emerging ordered state is typically time-periodic, a.k.a. a time crystal \cite{Sacha2018, Khemani2019}.

To illustrate the idea, let us first briefly review the concept of OBDP in the geometrically frustrated systems in equilibrium systems.
As we have seen, at zero temperature $T=0$ (no noise), geometrically frustrated systems often exhibit accidental degeneracy in their ground states (Fig.~\ref{fig: order by disorder schematic}(a),(c)). 
Mathematically, this can be described as the ground state energy $E_{\rm G}$ being independent of the system's configuration within the accidentally degenerate ground state manifold, which is parameterized by $\phi$ (i.e., $E_{\rm G}(\phi)={\rm const.}$).

Now, let us introduce thermal noise, corresponding to a system at finite temperature $T>0$ (Fig.~\ref{fig: order by disorder schematic}(b),(d)).
In this case, the system converges to a state that minimizes the free energy $F=E-TS$, where $S$ represents entropy. 
Although the energy $E=E_{\rm G}$ remains constant within the ground state manifold by definition, the fluctuation properties are typically configuration-dependent, resulting in a configuration-dependent entropy $S(\phi)$.
As a consequence, the accidental degeneracy is generically lifted entropically, driving the system towards the ground state with maximum entropy, $\phi_*$, which is ``selected'' (Fig.~\ref{fig: order by disorder schematic}(b)).
The selected state often exhibits a long-range order, giving rise to the counterintuitive phenomenon of OBDP, where thermal noise \textit{induces} order \cite{Villain1980, Henley1989, Moessner1998a, Moessner1998b,Bergman2007}.

In the language of dynamical systems, this can be translated into the dynamics of the (thermal averaged) parameter $\phi$ described by
\begin{eqnarray}
    \label{eq: entropic force geometrical}
    \dot\phi= -\frac{\partial F(\phi)}{\partial\phi}
    =-\frac{\partial E}{\partial \phi}
    +T\frac{\partial S(\phi)}{\partial\phi}
    =T\frac{\partial S(\phi)}{\partial\phi}.
\end{eqnarray}
The term $f_S \equiv T\partial S(\phi)/\partial \phi$ is an entropic force (or entropic torque, in the context of spin systems) induced by thermal noise, which drives the system towards the state of maximum entropy (i.e., the ``selected'' ground state, see Fig.~\ref{fig: order by disorder schematic}(c),(d)).
Crucially, the energy term $f_E\equiv-\partial E/\partial\phi$ vanishes because of the property that the system is marginal, which is what makes the entropic force $f_S\propto T$ dominant even at the weak thermal noise limit $T\rightarrow 0^+$. 
Note also that when the degeneracy originates from symmetry, the entropy $S$ cannot be configuration-dependent within the ground state degeneracy manifold because the symmetry guarantees their equivalence. 
Therefore,  in this case, $S(\phi)={\rm const.}$ and the entropic force is absent $f_S=0$.
This shows how the origin of degeneracy being \textit{accidental} is a key element for the emergence of OBDP.

We will demonstrate in this section that this concept can be generalized to non-reciprocally interacting systems (Fig.~\ref{fig: order by disorder schematic}(e),(f)). 
We will show that finite noise strength in a non-reciprocally interacting system can give rise to a conceptionally similar ``entropic force'' that drives a specific orbit towards stability, thereby triggering ``orbit selection'' among the ``accidentally degenerate'' orbits. 
In parallel to the case of geometrical frustration, we will demonstrate that this ``entropic force'' tends to favor an ordered phase, leading to noise-induced symmetry breaking --- a characteristic feature of OBDP. 
Importantly, with this concept extended to a broader class of dynamical systems, 
noise can now trigger phase transitions (bifurcations) beyond the conventional equilibrium paradigm.
We will showcase this by demonstrating a noise-induced non-reciprocal phase transition \cite{Fruchart2021} in our studied model, which has no counterpart in equilibrium systems.




\subsection{All-to-all coupled models}



To set the stage, we
consider an all-to-all coupled system where the spins are grouped into a few communities (labeled by $a,b=\text{A,B,C,...}$) that each consists of $N_a$ spins and are now subject to Gaussian white noise $\eta_i^a$,  
\begin{eqnarray}
    \label{eq: Langevin order-by-disorder}
    \dot\theta_i^a = 
    - \sum_b\frac{j_{ab}}{N_b} \sum_{j=1}^{N_b} \sin (\theta_i^a - \theta_j^b)
    + \eta_i^a,
\end{eqnarray}
where $\avg{\eta_i^a(t)}=0,\avg{\eta_i^a(t)\eta_j^b(t')}=\sigma\delta_{ab}\delta_{ij}\delta(t-t')$. 
We consider the case where the intra-community couplings are reciprocal and ferromagnetic $j_{aa}>0$, while the inter-community couplings may be non-reciprocal $j_{ab} \ne j_{ba}(a\ne b)$. 
This is an example of systems described by Eq.~\eqref{eq: many communities}.
The former causes the intra-community spins to order ferromagnetically at sufficiently weak noise strength, which is 
characterized by the order parameter $\psi_a(t) = (1/N_a) \sum_{i=1}^{N_a}e^{i\theta_i^a(t)}=r_a(t) e^{i\phi_a(t)}$ \cite{Kuramoto1984}.
Note that, for the reciprocal case $j_{ab}=j_{ba}$, this setup corresponds to an equilibrium system at finite temperature $T=\sigma/(2 k_{\rm B})$ (where $k_{\rm B}$ is the Boltzmann constant).


\begin{figure}[t]
\centering
\includegraphics[width=0.85\linewidth,keepaspectratio]{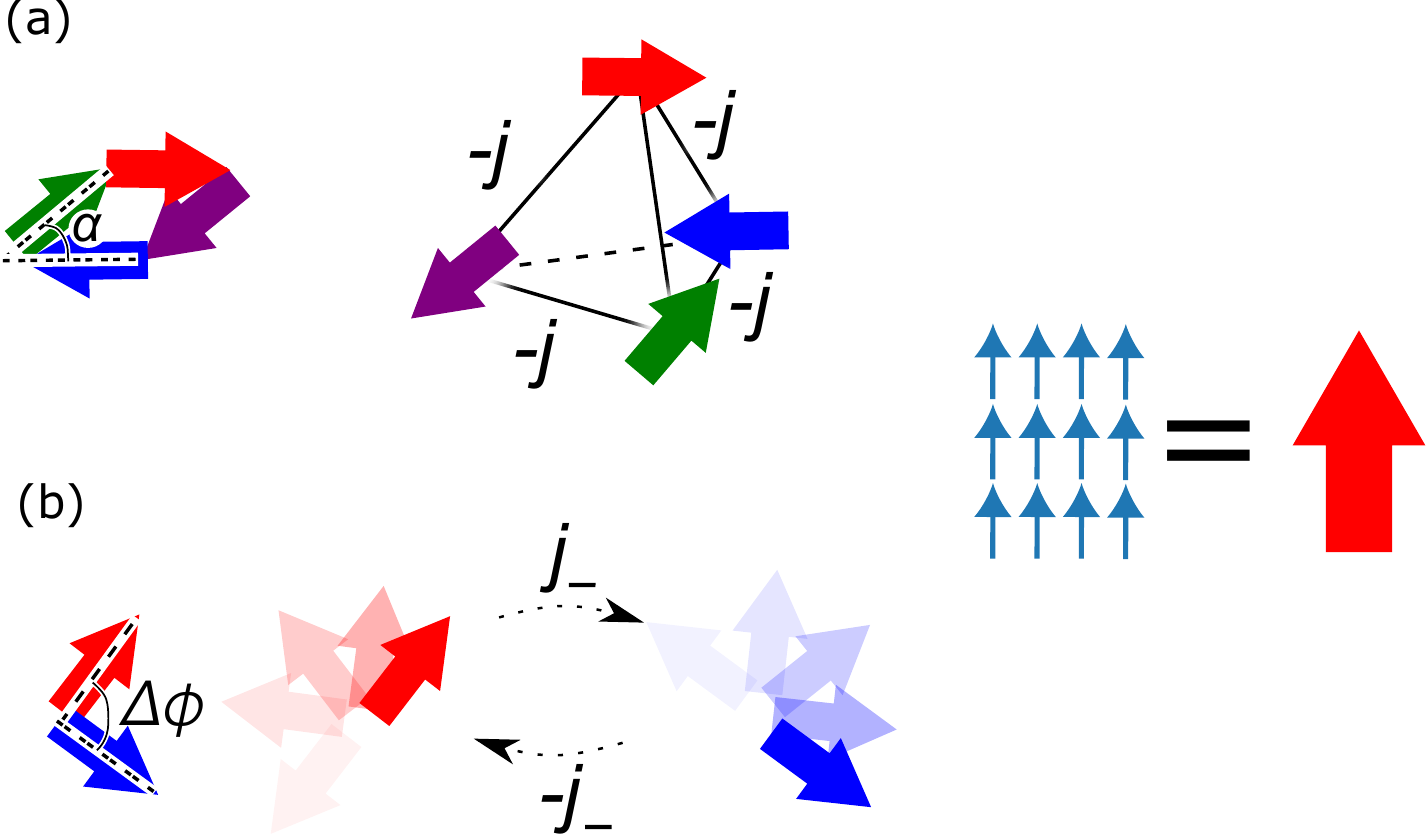}
\caption{
\textbf{``Accidental degeneracy'' of orbits in geometrically and non-reciprocally frustrated all-to-all coupled many-body systems. }
The thick arrows represent the macroscopic angles $\phi_a$ that are composed of a macroscopic number of spins represented by smaller solid arrows.
A geometrically frustrated four-community system illustrated in (a) exhibits an accidental degeneracy parameterized by a relative angle $\alpha$.
Similarly, a non-reciprocally frustrated two-community system illustrated in (b) exhibits marginal orbits parameterized by a relative angle $\Delta\phi$.
These degeneracies are shown to get lifted by introducing disorder to the system.}
\label{fig: order-by-disorder}
\end{figure}

In the absence of noise $\sigma=0$, as discussed in the final part of Sec.~\ref{sec: accidental degeneracy}, all of the spins in the same community would eventually align ($\theta_i^a=\phi_a$) to give perfect magnetization $r_a = 1$. 
As a result, the spins in the same community will collectively behave as a  macroscopic object 
that follows the same dynamics as Eq.~\eqref{eq: XY-model} (as we have discussed in  Eq.~\eqref{eq: phase order dynamics deterministic}).
Therefore, 
these macroscopic angles $\phi(t)=(\phi_{\rm A}(t),\phi_{\rm B}(t),\cdots)$ exhibit marginal, ``accidentally degenerate'' orbits 
when the inter-community couplings $j_{ab}$ are chosen to have geometrical or non-reciprocal frustration.
For example,  
in a geometrically frustrated system consisting of four communities $(a,b={\rm A,B,C,D})$ that  interacts antiferromagnetically $j_{ab}=-j<0(a\ne b)$ 
(Fig.~\ref{fig: order-by-disorder}(a)),
the system relaxes to the accidentally degenerate ground states parameterized by a relative angle $\alpha$ illustrated in the inset of Fig. \ref{fig: order-by-disorder}(a) (See Ref.~\cite{Moessner1998a, Moessner1998b} and Appendix \ref{Appendix: Order-by-disorder}).
Similarly, systems with non-reciprocal frustration with $j_{ab}=-j_{ba}$ exhibit time-dependent, marginal orbits $\phi(t)$. (See Fig.~\ref{fig: order-by-disorder}(b) and Fig.~\ref{fig: three spins}.)

Below, we show that this ``accidental degeneracy'' generically gets ``lifted" by the stochastic noise, irrespective of whether the degeneracy is originated from geometrical or non-reciprocal frustration.
In the presence of noise, $\theta_i^a$ fluctuates around the macroscopic spin angle $\phi_a$.  At sufficiently weak noise strength, the distribution of $\delta\theta_i^a=\theta_i^a-\phi_a$ takes a Gaussian distribution (See Appendix \ref{Appendix: Order-by-disorder}) 
\begin{eqnarray}
\label{eq: distribution function all-to-all}
    \rho_i^a(t,\delta\theta_i^a;\phi(t))
    =\frac{1}{\sqrt{\pi }w_a(t;\phi(t))}
    e^{-(\delta\theta_i^a )^2/w_a^2(t;\phi(t))}
\end{eqnarray}
with its width $w_a$ given by,
\begin{eqnarray}
    \label{eq: width}
    w_a^2(t;\phi(t))
    =2 \sigma \int _0^t d\tau e^{-2\int_{\tau}^t d\tau' 
    \sum_b j_{ab}
    \cos(\phi_a(\tau')-\phi_b(\tau'))}
    \nonumber
\end{eqnarray}
for an initial condition with a perfectly magnetized state $\delta\theta_i^a (t=0)= 0$. 
In many cases we will consider below, $\phi_a-\phi_b$ converges to a constant value in the long time limit, in which the width $w_a^2(t\rightarrow\infty,\phi)$ is given by,
\begin{eqnarray}
    w_a^2(t\rightarrow\infty,\phi)=\frac{\sigma}{\sum_b j_{ab}\cos(\phi_a - \phi_b)}.
\end{eqnarray}

Crucially, the width $w_a(\phi)$ of the fluctuations depends on which ``accidentally degenerate'' orbit $\phi(t)$ the system happened to take.
This is in stark contrast to the degenerate states arising from global symmetry, where all the degenerate states are guaranteed to have the same fluctuation properties by symmetry. 
The configuration-dependent fluctuation seen above is therefore a salient feature of the accidentally degenerate states.
Note how the ferromagnetic intra-community coupling $j_{aa}>0$ is playing a crucial role in preventing the width $w_a^2$ from becoming negative, ensuring the stability of the orbits. 

As a result, macroscopic angle dynamics
\begin{eqnarray}
    \label{eq: order parameter renormalized}
    \dot\phi_a(t) 
    = -\sum_b j_{ab}^\star(\phi(t))
    \sin(\phi_a(t)-\phi_b(t))
    +\bar\eta_a(t),
\end{eqnarray}
are now governed by $\phi$-\textit{dependent, renormalized} coupling 
\begin{eqnarray}
    \label{eq: renormalized coupling}
    &&j_{ab}^\star(\phi(t))
    = j_{ab}
    \frac{r_b(\phi(t))}{r_a(\phi(t))}
    \avg{
    \cos^2\delta\theta_i^a
    }_{\phi(t)},
\end{eqnarray}
where $\avg{h(\delta\theta_i^a)}_{\phi(t)}=\int d\theta_i^a \rho_i^a(t,\delta\theta_i^a;\phi(t))h(\delta\theta_i^a)$  (See Appendix \ref{Appendix: Order-by-disorder} for derivation). 
Here, we have assumed that the system self-averages, $\avg{h(\delta\theta_i^a)}_{\phi(t)} = (1/N_a)\sum_{i=1}^{N_a}h(\delta\theta_i^a)$.
In Eq.~\eqref{eq: renormalized coupling}, $\bar\eta_a\approx(1/N_a)\sum_{i=1}^{N_a}\eta_i^a$ is the noise acting on the macroscopic angle $\phi_a$ that obeys $\avg{\bar\eta_a(t)}=0$ and 
\begin{eqnarray}
    \label{eq: macroscopic noise strength}
    \avg{\bar\eta_a(t)\bar\eta_b(t')}\approx\frac{\sigma}{N_a}
    \delta_{ab}\delta(t-t').    
\end{eqnarray}
As a result of $\phi_a(t)$ being a  macroscopic quantity, the noise strength on this quantity vanishes as one takes the thermodynamic limit $N_a\rightarrow \infty$.

The renormalization of the coupling gives rise to an additional torque to the deterministic limit (Eq.~\eqref{eq: phase order dynamics deterministic}), which can be regarded as the entropic torque generalized to dynamical systems that are not necessarily written in terms of free energy (\textit{cf.},  Eq.~\eqref{eq: entropic force geometrical}).
As we will see, this entropic contribution determines the macroscopic features of systems with geometrical or non-reciprocal frustration that have marginal orbits.


First consider the geometrically frustrated system introduced above, which consists of four communities that antiferromagnetically interact ($j_{ab}=-j<0(a\ne b)$).
This system has an accidentally degenerate ground state manifold parameterized by an angle $\alpha$  (Fig. \ref{fig: order-by-disorder}(a)).
%
In this situation, the effective coupling turns out to be $\phi$-independent $j_{ab}^\star(\phi(\alpha))=-j^\star<0$ on this manifold (Appendix \ref{Appendix: Order-by-disorder}). 
Therefore, this many-body problem maps to that of a four-spin system on a tetrahedron lattice, but importantly, 
at a very low but \textit{finite} temperature $T\sim \sigma/N_a\rightarrow 0^+>0$.
As pointed out in Ref. \cite{Moessner1998b}, under such stochasticity, 
the probability to realize the angle $\alpha$ is given by the Boltzmann distribution (where $F(\alpha)$ is a free energy and $S(\alpha)$ is the entropy at configuration $\alpha$) (See Appendix \ref{Appendix: Order-by-disorder} Sec. \ref{Appendix subsubsection: geometrically frustrated case} for derivation),
\begin{eqnarray}
\rho(\alpha)\propto
e^{-F(\alpha)/(k_{\rm B}T)}=e^{S(\alpha)/k_{\rm B}}\sim|\sin(\alpha)|^{-1},
\label{eq: tetrahedron order by disorder}
\end{eqnarray}
for $\sin^2\alpha\gg \sigma/(N_a j^\star)\rightarrow 0$ 
that 
is found to be overwhelmingly concentrated to  the collinear configuration $\alpha_*=0,\pi$.
In other words, the entropic effects ``select'' the collinear configuration $\alpha_*=0,\pi$ among the degenerate ground states (or the marginal fixed points in the dynamical system language), giving rise to an OBDP.


\begin{figure}[t]
\centering
\includegraphics[width=0.7\linewidth,keepaspectratio]{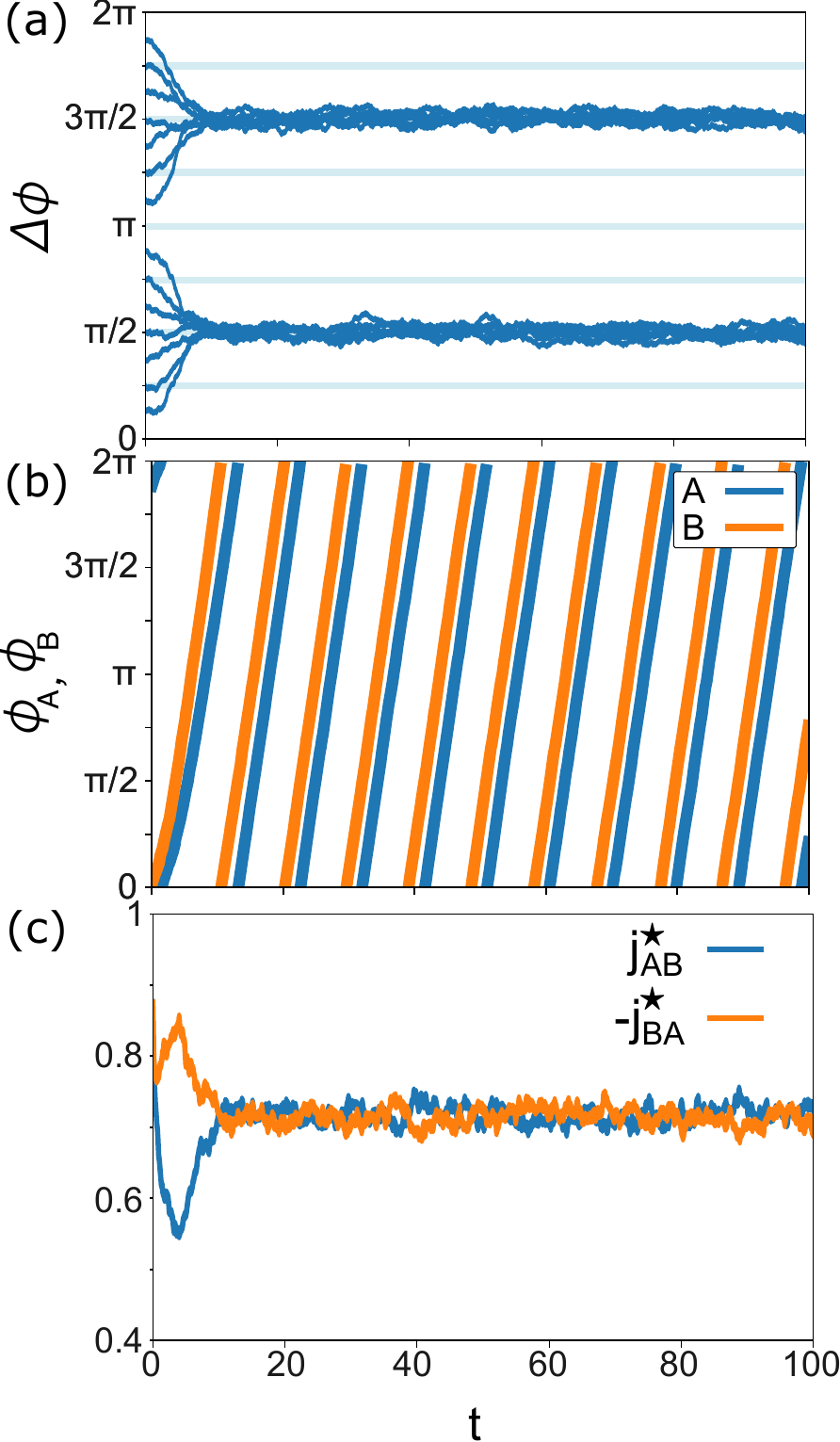}
\caption{
\textbf{Time crystalline order-by-disorder phenomena induced by non-reciprocal frustration.}
Time evolution of (a) the angle difference $\Delta\phi=\phi_{\rm A}-\phi_{\rm B}$,
(b) $\phi_{\rm A},\phi_{\rm B}$, and
(c) the effective coupling strength $j_{ab}^\star(\phi)$ (Eq.~\eqref{eq:  renormalized coupling}). 
In (a), the solid (thin) line represents the dynamics in the presence (absence) of noise.  
We set $j_{\rm AA}=j_{\rm BB}=3,j_{\rm AB}=-j_{\rm BA}=1$, the noise strength $\sigma=1.5$, and the number of spins $N_{\rm A}=N_{\rm B}=2000$.
The system ``selects'' $\Delta\phi_*=\pm\pi/2$ that satisfies 
Eq.~\eqref{eq:  chiral phase condition}
to give rise to the time-dependent phase (chiral phase), all in agreement with our analytical analysis in the main text.
}
\label{fig: order-by-disorder_result}
\end{figure}

We show below that a similar ``orbit selection'' takes place in non-reciprocally frustrated systems as well (Fig.~\ref{fig: order-by-disorder}(b)), due to the entropic force $f_S$ that is analogous to those arising in Eq.~\eqref{eq: entropic force geometrical}.
To be explicit, let us consider the case of two communities $a={\rm A,B}$ that are non-reciprocally coupled ($j_{\rm AB}\ne j_{\rm BA}$). 
In the deterministic case $\sigma=0$ , since the dynamics of the order parameter is given by Eq.~\eqref{eq:  phase order dynamics deterministic},
the angle difference $\Delta\phi=\phi_{\rm A}-\phi_{\rm B}$ and the center-of-mass angle $\Phi = (\phi_{\rm A}+\phi_{\rm B})/2$ dynamics is governed by
\begin{eqnarray}
    \label{eq:  deterministic two collective spin Delta phi}
    \Delta\dot\phi 
    = -(j_{\rm AB}+j_{\rm BA})
    \sin\Delta\phi, \qquad &(\sigma=0)&
    \\
    \label{eq:  deterministic two collective spin Phi}
    \dot\Phi=-\frac{j_{\rm AB}
    -j_{\rm BA}}{2}\sin\Delta\phi.
    \qquad&(\sigma=0)&
\end{eqnarray}
Hence, for the perfectly non-reciprocal case $j_{\rm AB}=-j_{\rm BA}=j_-$ where the Liouville-type theorem (Eq.~\eqref{eq:  Liouville's theorem}) holds,
the angle difference is initial state dependent $\Delta\phi(t)=\Delta\phi(0)$. 
The ``accidentally degenerate orbits'' is parameterized by $\Delta\phi$ in this case (Fig.~\ref{fig: order-by-disorder}(b)).



We will now show that the ``orbit selection'' occurs in the presence of noise $\sigma>0$ due to the emergence of entropic torque.
From Eq.~\eqref{eq:  order parameter renormalized},
$\Delta\phi$ and $\Phi$ dynamics
is governed by the equation of motion determined by the renormalized coupling $j_{ab}^\star(\Delta\phi)$ (Eq.~\eqref{eq:  renormalized coupling}),
\begin{eqnarray}
    \label{eq:  phase order dynamics renormalized}
    \Delta\dot\phi
    &=&-(j_{\rm AB}^\star(\Delta\phi)+j_{\rm BA}^\star(\Delta\phi))
    \sin\Delta\phi,
    \\
    \label{eq:  phase order dynamics renormalized center-of-mass}
    \dot\Phi &=& -\frac{j_{\rm AB}^\star(\Delta\phi)-j_{\rm BA}^\star(\Delta\phi)}{2}
    \sin\Delta\phi.
\end{eqnarray}
Here, we have dropped the macroscopic noise $\bar\eta_a(t)$, which is justified in the thermodynamic limit $N_a\rightarrow\infty$. (See Eq.~\eqref{eq:  macroscopic noise strength}.)
Due to the $\Delta\phi$-dependence of the renormalized couplings $j_{ab}^\star(\Delta\phi)$, 
the Liouville-type theorem no longer holds and
the angle difference $\Delta\phi$ exhibits \textit{stable} fixed points even in the non-reciprocal limit $j_{\rm AB}=-j_{\rm BA}$; the ``orbit selection'' occurs.
In particular, there are two candidates for stable fixed points of $\Delta\phi$ (Eq.~\eqref{eq:  phase order dynamics renormalized}) that correspond to different phases of matter.
One is a phase that satisfies         
\begin{eqnarray}
    \label{eq:  static phase condition}
    \sin\Delta\phi_*=0,
\end{eqnarray}
which corresponds to a static phase $\dot\Phi=0$ that has an aligned ($\Delta\phi_*=0$) or an anti-aligned ($\Delta\phi_*=\pi$) configuration.

The other is a phase that only emerges in the presence of noise $\sigma>0$, which satisfies
\begin{eqnarray}
    \label{eq:  chiral phase condition}
    j_{\rm AB}^\star(\Delta\phi_*) = - j_{\rm BA}^\star(\Delta\phi_*).
\end{eqnarray}
Generically, $\Delta\phi_*\ne 0,\pi$, corresponding to a time-dependent phase $\dot\Phi \ne 0$ that is referred to as a chiral phase in Ref.~\cite{Fruchart2021}.
Importantly,  while the static phase is invariant under the parity operation, 
\begin{eqnarray}
    \label{eq:  parity}
    (\phi_{\rm A},\phi_{\rm B})\rightarrow (-\phi_{\rm A},-\phi_{\rm B}),
\end{eqnarray} 
the chiral phase spontaneously breaks it.
Note that the renormalized coupling satisfies $j_{ab}^\star(\Delta\phi)=j_{ab}^\star(-\Delta\phi)$, which follows from the property  $\rho_i^a(\delta\theta_i^a;\Delta\phi)=\rho_i^a(\delta\theta_i^a;-\Delta\phi)$ (that can be obtained from Eqs.~\eqref{eq:  distribution function all-to-all} and \eqref{eq:  renormalized coupling}). 
Therefore, if $\Delta\phi=\Delta\phi_*$ were found to be a stable fixed point of Eq.~\eqref{eq:  phase order dynamics renormalized}, then $\Delta\phi=-\Delta\phi_*$ must also be a stable fixed point, which transforms one to the other via parity operation \eqref{eq:  parity}.


\begin{figure}[t]
\centering
\includegraphics[width=0.9\linewidth,keepaspectratio]{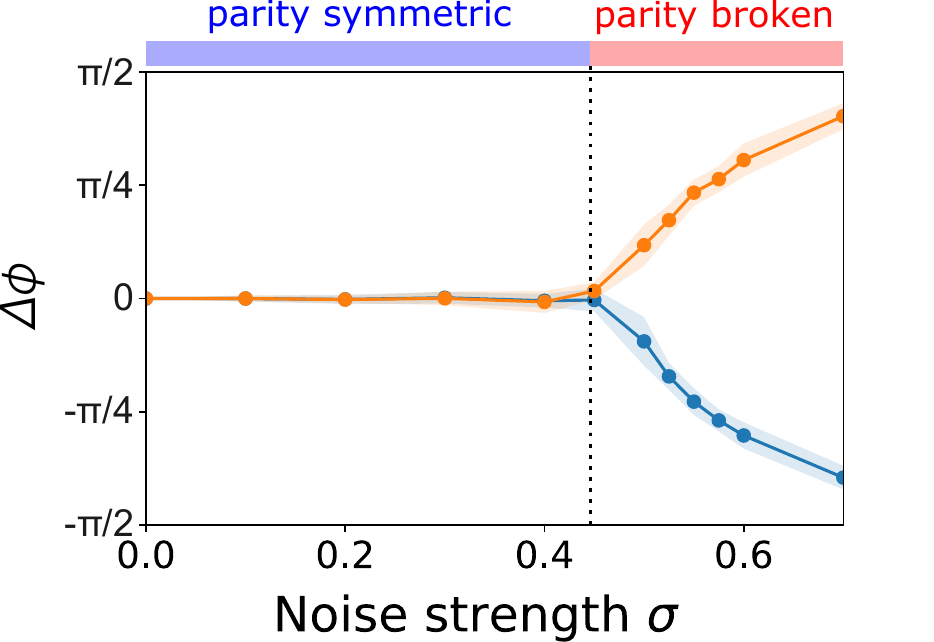}
\caption{
\textbf{Noise induced spontaneous parity breaking.}
The computed noise strength dependence of the angle difference $\Delta\phi=\phi_{\rm A}-\phi_{\rm B}$ of the order parameter $\psi_a=r_a e^{i\phi_a}$ in the steady state for $j_{\rm AB}=0.35\ne -j_{\rm BA}=0.25$.
Here, the solid lines and the shaded area represent the average and the variance of $\Delta\phi$, respectively, which were calculated using the data in the time range $100<t<400$ with an initial condition $\theta_i^{\rm A}(t=0)=\pi/4 (=-\pi/4)$ and $\theta_i^{\rm B}(t=0)=0$ for blue (orange) plots.
The transition from $\Delta\phi=0$ (parity symmetric static phase) to $\Delta\phi\ne 0$ (parity broken chiral phase) as $\sigma$ increases indicates that the noise has induced the spontaneous symmetry breaking, which is a salient feature of OBDP
and is opposite from what is usually expected.
We set $j_{\rm AA}=j_{\rm BB}=1$, and
$N_{\rm A}=N_{\rm B}=2000$.
}
\label{fig: z2ssb}
\end{figure}

\begin{figure}[t]
\centering
\includegraphics[width=0.95\linewidth,keepaspectratio]{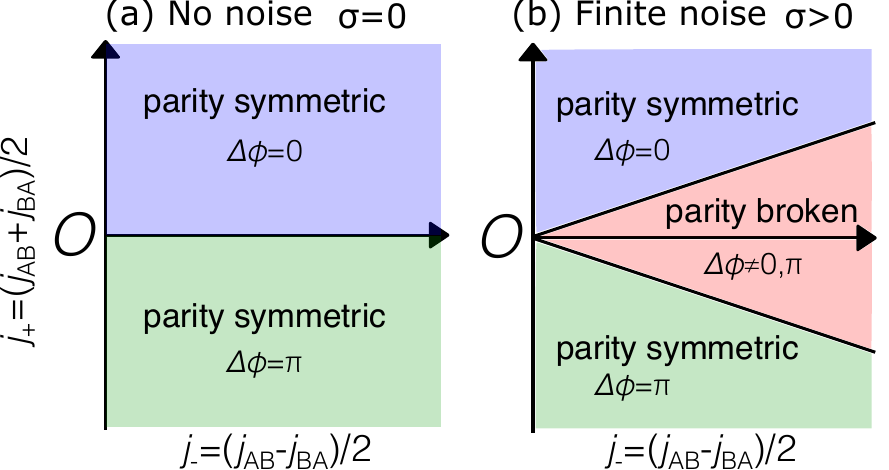}
\caption{
\textbf{Schematic phase diagram.}
(a), (b) Schematic steady-state phase diagram in the absence of noise $\sigma=0$ (a) and in the presence of noise $\sigma>0$ (b). 
The intra-community couplings $j_{\rm AA}>0$ and $j_{\rm BB}>0$ are assumed to be large compared to inter-community couplings $j_{\rm AB}$ and $j_{\rm BA}$.}
\label{fig: phase diagram}
\end{figure}

Take a perfectly non-reciprocal system $j_{\rm AB}=-j_{\rm BA}=j_-$ that has an identical intra-community ferromagnetic coupling strength between the two communities $j_{\rm AA}=j_{\rm BB}=j_0$ (taken to be $j_0>|j_-|$ to ensure stability $w_a^2>0$) as an example. In this case, Eq.~\eqref{eq:  phase order dynamics renormalized} reads (See Appendix \ref{Appendix: Order-by-disorder})
\begin{eqnarray}
\label{eq:  Delta phi jAA=jBB}
    \Delta\dot\phi
    &\simeq&
    \frac{j_0 j_-^2\sigma^2}{2}
    \frac{ \cos\Delta\phi }{(j_0^2 - j_-^2 \cos^2\Delta\phi)^2}
    \sin\Delta\phi
\end{eqnarray}
at sufficiently weak noise level, which has stable fixed points at $\Delta\phi_* = \pm\pi/2$ that corresponds to a chiral phase, satisfying Eq.~\eqref{eq:  chiral phase condition}.
The fixed points $\Delta\phi_*=0,\pi$ are unstable. 
All these features are consistent with the numerical result presented in Fig.~\ref{fig: order-by-disorder_result}.
This clearly shows that the noise $\sigma>0$ has induced an ``entropic torque'' that stabilizes a spontaneous parity broken phase, in a similar manner to the geometrically frustrated case, \textit{cf.}, Eq.~\eqref{eq:  entropic force geometrical}.
Derivation of this ``entropic torque'' induced by non-reciprocity is one of the main results of this paper.

\begin{figure}[t]
\centering
\includegraphics[width=0.6\linewidth,keepaspectratio]{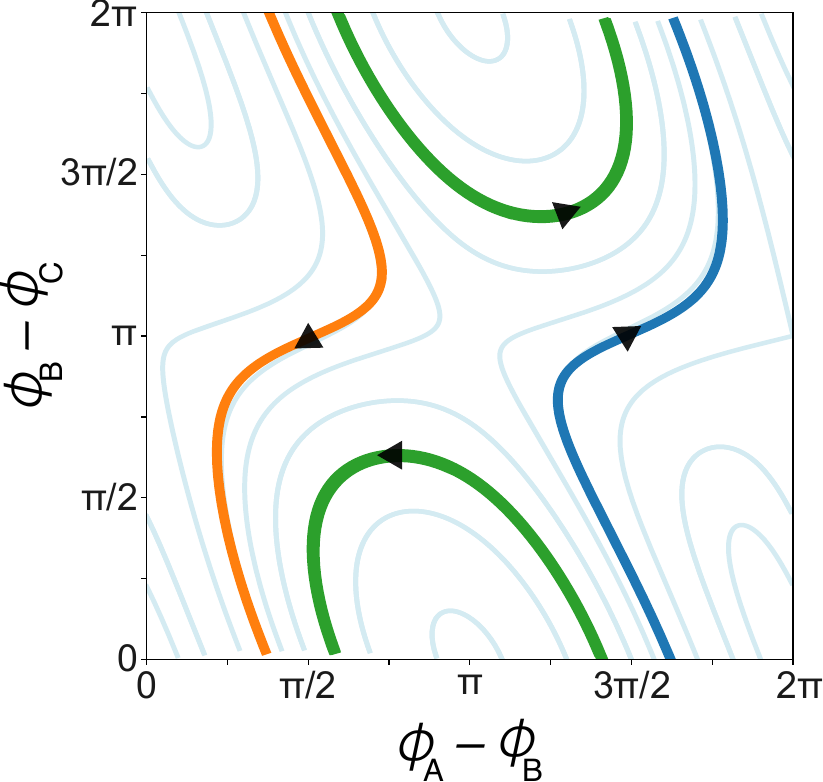}
\caption{
\textbf{Order-by-disorder phenomena in a non-reciprocally coupled three community system with random torque.}
Solid (thin) lines represent the trajectories for different initial conditions in the presence (absence) of random torque, where one can see that certain orbits are ``selected'' by the disorder.
The angle difference $\phi_a-\phi_b$ dynamics  are computed using the Ott-Antonsen ansatz \cite{Ott2008,Ott2009} (Eq.~\eqref{SIeq:  Ott-Antonsen} in Appendix \ref{Appendix: Order-by-disorder}).
We set the coupling $j_{\rm AA}=j_{\rm BB}=j_{\rm CC}=4,j_{\rm AB}=-j_{\rm BA}=3,j_{\rm BC}=-j_{\rm CB}=-1,j_{\rm CA}=-j_{\rm AC}=2$, the torque distribution width $\Delta=0.1$.
}
\label{fig: three community case}
\end{figure}


This property has an important implication to a more general case, i.e. when one is away from the perfectly non-reciprocal case $j_{\rm AB}\ne -j_{\rm BA}$.
Figure \ref{fig: z2ssb} shows the noise strength dependence of $\Delta\phi_*$ when $j_{\rm AB}=0.35\ne -j_{\rm BA}=0.25$.
While at small noise strength $\sigma$, the parity-symmetric static phase $\Delta\phi_*=0$ is realized, a parity-broken chiral phase $\Delta\phi_*\ne 0$ emerges at \textit{higher} noise level.
This noise-induced spontaneous symmetry breaking is a salient feature of OBDP.

A qualitative understanding of this counter-intuitive phenomenon can be obtained from our formalism as follows.
In the deterministic case, when the reciprocal part of the inter-community coupling is positive (negative) $j_+\equiv(j_{\rm AB}+j_{\rm BA})/2>0(<0)$, Eq.~\eqref{eq:  deterministic two collective spin Delta phi} tells us that a static phase with an aligned configuration  $\Delta\phi_*=0(\pi)$ would be realized, in agreement with Fig.~\ref{fig: z2ssb} at $\sigma=0$.
As one turn on the noise strength $\sigma>0$, Eq.~\eqref{eq:  phase order dynamics renormalized} reads,
\begin{eqnarray}
    \label{eq:  phase order dynamics renormalized small j+}
    \Delta\dot\phi
    = \bigg[
    -2 j_+ 
    +\frac{j_0 j_-^2\sigma^2}{2}
    \frac{ \cos\Delta\phi }{(j_0^2 - j_-^2 \cos^2\Delta\phi)^2}
    \bigg]
    \sin\Delta\phi,
    \nonumber\\
\end{eqnarray}
where we have assumed small reciprocity $|j_+|\ll |j_-|,j_0$ and restricted ourselves to be near the phase transition point. 
(See Appendix \ref{Appendix: Order-by-disorder} for derivation.) 
This equation is to be compared with its geometrical frustration counterpart, Eq.~\eqref{eq: entropic force geometrical}.
Here, the first term proportional to the reciprocal piece $j_+>0(<0)$ describes the torque that tries to make the angles (anti-)aligned, which can be considered as an analog of the first term of Eq.~\eqref{eq: entropic force geometrical}.
This torque competes with the ``entropic torque'' induced by non-reciprocal frustration (the second term that is identical to the right-hand side of Eq.~\eqref{eq:  Delta phi jAA=jBB}).

When the noise is weak and the first term is dominant over the second, Eq.~\eqref{eq:  phase order dynamics renormalized small j+} only has one stable fixed point $\Delta\phi_*=0(\pi)$.
However, once the noise strength $\sigma$ exceeds a critical value,  
\begin{eqnarray}
    \label{eq:  phase order dynamics critical}
    \sigma > \sigma_c
    =2\sqrt{\frac{|j_+|}
    {j_0 }}
    \frac{j_0^2 - j_-^2}{|j_-|}
\end{eqnarray}
the entropic torque (the second term of Eq.~\eqref{eq:  phase order dynamics renormalized small j+}) makes the system bifurcate to a chiral phase with $\phi_*\ne 0,\pi$.
This signals the emergence of a spontaneous parity breaking seen in Fig.~\ref{fig: z2ssb}. This result implies a phase diagram schematically depicted in Fig.~\ref{fig: phase diagram}.
We remark that the phase transition observed above is an instance of a non-reciprocal phase transition \cite{Fruchart2021},
which has no equilibrium counterpart.
Non-reciprocal phase transition is marked by a spectral singularity called an exceptional point, where the eigenvectors coalesce \cite{Kato1984} to a zero mode at the critical point \cite{Hanai2020, Fruchart2021}.
This can be seen by linearizing Eqs.~\eqref{eq:  phase order dynamics renormalized small j+} and \eqref{eq:  phase order dynamics renormalized center-of-mass} around the static phase $\Delta\phi=\Delta\phi_*+\delta\Delta\phi=\delta\Delta\phi$,
\begin{eqnarray}
    \begin{pmatrix}
        \delta\dot\Phi  \\
        \delta\Delta\dot\phi
    \end{pmatrix}
    &=&
    \begin{pmatrix}
        0 & -2 j_-
        + \frac{j_0\sigma}{j_0^2-j_-^2} + \frac{(2j_0^2+j_-^2)\sigma^2}{4(j_0^2-j_-^2)^2}
        \\
        0 & -2 j_+ + \frac{j_0 j_-^2\sigma^2}{2(j_0^2-j_-^2)} 
    \end{pmatrix}
    \begin{pmatrix}
        \delta\Phi  \\
        \delta\Delta\phi
    \end{pmatrix}
    \nonumber\\
    &=&
    \hat L
    \begin{pmatrix}
        \delta\Phi  \\
        \delta\Delta\phi
    \end{pmatrix}.
\end{eqnarray}
At the transition point $\sigma=\sigma_c$ (Eq.~\eqref{eq:  phase order dynamics critical}), the (2,2)-component of the Jacobian $\hat L$ vanishes, giving a defective matrix \cite{Kato1984} with zero eigenvalues, 
\begin{eqnarray}
    \hat L_c = 
    \begin{pmatrix}
        0 & -2j_-
       +2\sqrt{j_-j_+} -\frac{j_+j_-}{j_0}-\frac{2j_0j_+}{j_-}
        \\
        0 & 0
    \end{pmatrix}.
\end{eqnarray}
This shows that the eigenvectors with zero eigenvalue coalesce at the critical point, which is a salient feature of  non-reciprocal phase transitions~\cite{Fruchart2021,Hanai2019,Hanai2020,Saha2020,You2020,Zelle2023}.



It is also worth mentioning that there are cases where the ``entropic effects'' favor a static state, even in the perfectly non-reciprocal case $j_+=0$. 
In Appendix \ref{Appendix: Order-by-disorder}, we show both analytically and numerically that systems with $j_{\rm AA}\gg j_{\rm BB}$ and $j_{\rm AA}\ll j_{\rm BB}$ ``select'' a static state, $\Delta\phi_*=\pi$ and $\Delta\phi_*=0$, respectively (See Fig. \ref{Extended figure: different jaa} in Appendix \ref{Appendix: Order-by-disorder}).
Physically, this is due to the property that, when $j_{\rm AA}\gg j_{\rm BB} (j_{\rm AA}\ll j_{\rm BB})$,  width of the fluctuation of the A(B) community $w_{\rm A}(w_{\rm B})$ is smaller because the A(B) community gets stiff, leading to stronger suppression of $|j_{\rm AB}^\star(\Delta\phi)|(|j_{\rm BA}^\star(\Delta\phi)| )$. (See Eq.~\eqref{SIeq: renormalized coupling approx}.)
Similar effects can be seen when the noise strengths are different between the two communities, i.e., when the noise is characterized by $\avg{\eta_a(t)}=0,\avg{\eta_a(t)\eta_b(t')}=\sigma_{ab}\delta_{ab}\delta(t-t')$ with $\sigma_{\rm A}\ne\sigma_{\rm B}$. 
When $\sigma_{\rm A}\gg\sigma_{\rm B}(\ll \sigma_{\rm B})$ with $j_{\rm AA}=j_{\rm BB}$ leads to ``selecting'' the (anti-)aligned static configuration $\Delta\phi_*=0(\pi)$. 
These illustrate how the ``entropic torques'' that determine which state the system selects are strongly affected by the fluctuation properties of the degenerate states.

\begin{figure}[t]
\centering
\includegraphics[width=0.6\linewidth,keepaspectratio]{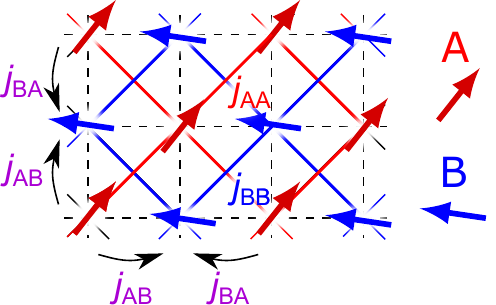}
\caption{
\textbf{Non-reciprocal XY-spin system on a hypercubic lattice.}
In this model, XY spins on a hypercubic lattice interact with their next-nearest-neighbor (nearest-neighbor) spins in a reciprocal (non-reciprocal) manner. The spins are divided into two communities, namely the sublattices A and B. 
Within each sublattice, the spins interact reciprocally via next-nearest-neighbor interactions $j_{\rm AA}$ and $j_{\rm BB}$, while the spins on different sublattices interact non-reciprocally through nearest-neighbor interactions $j_{\rm AB}$ and $j_{\rm BA}$. The figure illustrates the case of a two-dimensional spatial dimension ($d = 2$), but the model can be readily extended to higher spatial dimensions.
}
\label{fig: model spatial}
\end{figure}

So far, we have analyzed the simplest system with two communities under stochastic noise.
However, the underlying mechanism of OBDP is not restricted to such a specific case. 
We show below that an orbit selection occurs for 
a non-reciprocal \textit{three}-community system 
\begin{eqnarray}
    \label{eq:  Kuramoto}
    \dot\theta_i^a = \omega_i^a - 
    \sum_b \frac{j_{ab}}{N_b}
    \sum_{j=1}^{N_b}\sin(\theta_i^a-\theta_j^b).
\end{eqnarray}
with a random torque $\omega_i^a$ ($a,b={\rm A,B,C}$) distributed in a Lorentz distribution function
\begin{eqnarray}
    p_a(\omega_i^a)=\frac{1}{\pi}
    \frac{\Delta}{(\omega_i^a)^2+\Delta^2}
\end{eqnarray}
as a source of quenched disorder.
The width $\Delta$ characterizes the strength of the quenched disorder.
This is the Kuramoto model \cite{Kuramoto1984,Acebron2005,Fruchart2021} generalized to multiple communities.
According to Refs.~\cite{Ott2008,Ott2009},  the order parameter dynamics of this system are governed by,
\begin{eqnarray}
\label{SIeq:  Ott-Antonsen}
    \dot\psi_a = - \Delta\psi_a
    +\frac{1}{2}\sum_b j_{ab}(\psi_b - \psi_a^2 \psi_b^*). 
\end{eqnarray}
Figure \ref{fig: three community case} shows the order parameter dynamics of this system.
Among the marginal orbits in the absence of disorder $\Delta=0$ shown in  Fig.~\ref{fig: three spins} (and the thin line of Fig.~\ref{fig: three community case},
certain orbits are ``selected'' to be stable (solid lines in Fig.~\ref{fig: three community case}), signaling the occurrence of OBDP.



\subsection{Spatially extended models}

Systems that  exhibit the non-reciprocal frustration-induced OBDP is not restricted to all-to-all coupled models analyzed in the previous section. 
To demonstrate the generality of our finding, we now 
consider a non-reciprocal XY-spin system on a $d$-dimensional hypercubic lattice with nearest and next-nearest neighbor interactions. ($d=2$ case is  
illustrated in Fig. \ref{fig: model spatial}.)
Like in the model considered in the previous section, this model is composed of two communities of spins on different sublattices $a={\rm A,B}$. 
While the spins on the same sublattice interact ferromagnetically via the next-nearest neighbor interaction $j_{\rm AA},j_{\rm BB}>0$,
the spins on different sublattices interact non-reciprocally by the nearest-neighbor interaction $j_{\rm AB}\ne j_{\rm BA}$.
This system can be regarded as a natural extension of the all-to-all coupled model studied in the previous section generalized to a spatially extended $d$-dimensional system.
The stochastic equation of motion is given by,
\begin{eqnarray}
\label{eq: Langevin d-dim A}
\dot\theta_{\bm x}^{\rm A} & = &
\frac{1}{2d}
\sum_{\hat a}
j_{\rm AA}\sin(\theta_{\bm x+\hat a}^{\rm A} - \theta_{\bm x}^{\rm A})
\nonumber\\
&+&
\frac{1}{2d}
\sum_{\hat b}
j_{\rm AB}\sin(\theta_{\bm x+\hat b}^{\rm B} - \theta_{\bm x}^{\rm A})
+\eta_{\bm x}^{\rm A},
\\
\label{eq: Langevin d-dim B}
\dot\theta_{\bm x}^{\rm B} & = &
\frac{1}{2d}
\sum_{\hat a}
j_{\rm BB}\sin(\theta_{\bm x+\hat a}^{\rm B} - \theta_{\bm x}^{\rm B})
\nonumber\\
&+&\frac{1}{2d}
\sum_{\hat b}
j_{\rm BA}\sin(\theta_{\bm x+\hat b}^{\rm A} - \theta_{\bm x}^{\rm B})
+\eta_{\bm x}^{\rm B}.
\end{eqnarray}
Here, the angle of spin in the position $\bm x$ on a sublattice ${\rm A(B)}$ is denoted by $\theta_{\bm x}^{\rm A(B)}$ and
$\hat a$ and $\hat b$ are the vectors that point to the next-nearest and nearest neighbor sites, respectively. $\eta_{\bm x}^{a}(t)$ is a white noise characterized by $\avg{\eta_{\bm x}^{a}(t)}=0$ and $\avg{\eta_{\bm x}^a(t)\eta_{\bm x'}^b(t')}=\sigma \delta_{ab}\delta_{\bm x,\bm x'}\delta(t-t')$.




\begin{figure}[b]
\centering
\includegraphics[width=0.9\linewidth,keepaspectratio]{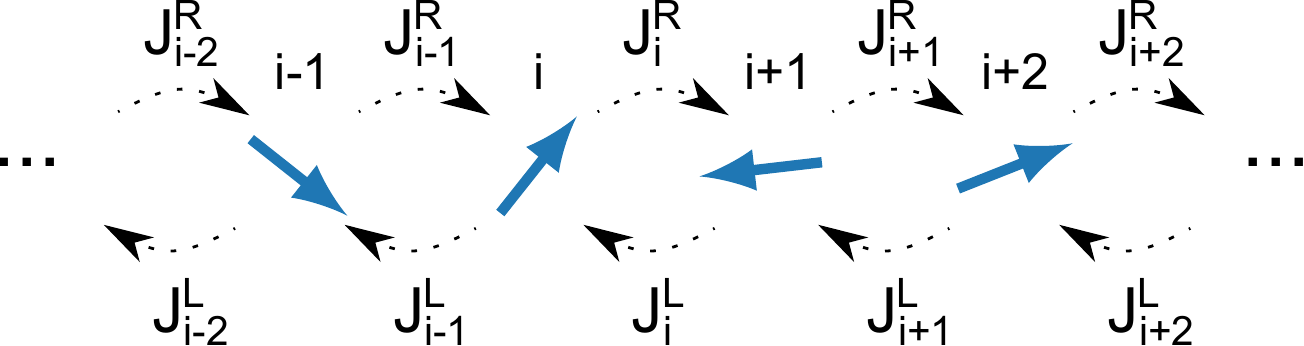}
\caption{
\textbf{One-dimensional spin chain with random asymmetric nearest-neighbor coupling.}
}
\label{fig: glass model}
\end{figure}

\begin{figure*}[t]
\centering
\includegraphics[width=0.96\linewidth,keepaspectratio]{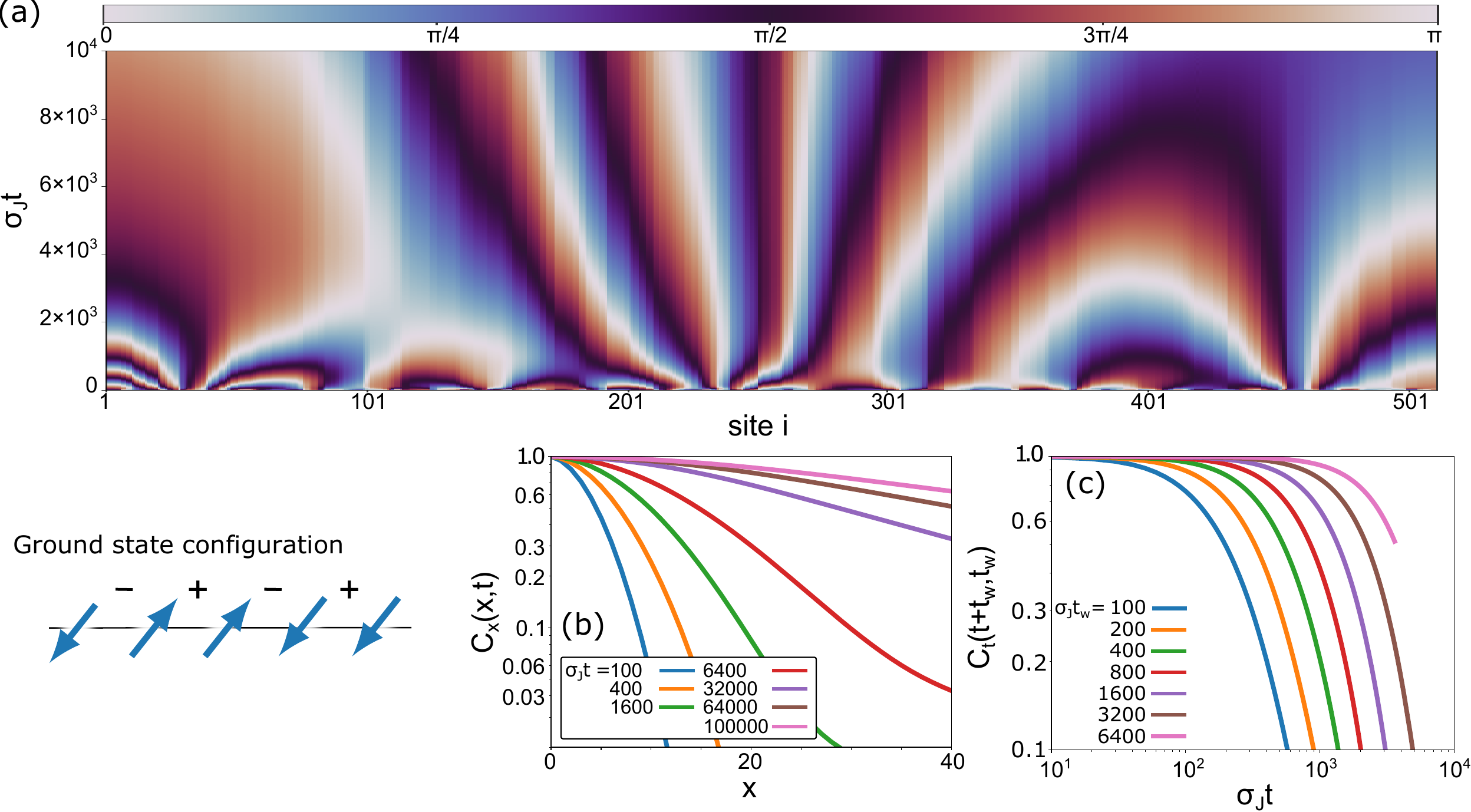}
\caption{
\textbf{Domain wall annihilation dynamics in reciprocal one-dimensional random spin chain.}
The reciprocal coupling $J_i^{\rm R}=J_i^{\rm L}$ ($J_{ij}=J_{ji}$) case.
As shown in the bottom-left panel, the ground state configuration of the reciprocally coupled spin chain (where the signs represent the sign of the reciprocal coupling at each bond) exhibits nematic order that is unique up to global rotation, implying the absence of geometrical frustration. 
(a) Typical trajectory of (nematic) angles $\varphi_i=\theta_i({\rm mod}~\pi)$. 
We set $N=2^9=512$ and the initial conditions were taken randomly from a uniform distribution $\theta_i=[0,2\pi)$.
(b) Spatial correlation function $C_x(x,t)$. 
(c) Time correlation function $C_t(t_w+t,t_w)$. In panels (b) and (c), we have averaged over $400$ trajectories of random initial conditions and configurations of coupling strengths and have set $N=2^{10}=1024$.
The domain wall annihilation dynamics of this one-dimensional chain give rise to slow relaxation (that shows aging phenomena) towards a long-ranged nematically ordered state.}
\label{fig: glass reciprocal}
\end{figure*}

In parallel to the previous section,
we first consider the deterministic case, $\sigma=0$.
In this case, as before, 
there is nothing that disturbs the spins from aligning within the same sublattice.
As a result, all the angles in the same sublattices align,
$\theta_{\bm x}^{\rm A}=\phi_{\rm A}$ and $\theta_{\bm x}^{\rm B}=\phi_{\rm B}$, where, as before, we have introduced the order parameter $\psi_a = r_a e^{i\phi_a}=N^{-1}\sum_{\bm x}e^{i\theta_{\bm x}^a}$ (where $N$ is the number of spins at each sublattice).
In such a case, 
Eqs.~\eqref{eq: Langevin d-dim A} and \eqref{eq: Langevin d-dim B} reduce to the same dynamics as the two spin system
(Eq. \eqref{eq:  phase order dynamics deterministic}).
Therefore, in the case of perfect non-reciprocity $j_{\rm AB}=-j_{\rm BA}$, 
the ``accidental degeneracy'' of orbits arises, parameterized again by the relative angle $\Delta\phi=\phi_{\rm A}-\phi_{\rm B}$.

In the presence of the noise $\sigma>0$, OBDP occurs.
Due to the property that the distribution of fluctuations around the steady state $\delta\theta_{\bm x}^a=\theta_{\bm x}^a-\phi_a$ is strongly dependent on the orbit of the order parameter $\Delta\phi$,
the order parameter dynamics follows the same form as Eqs.~\eqref{eq:  phase order dynamics renormalized} and \eqref{eq:  phase order dynamics renormalized center-of-mass}, with the renormalized coupling 
that has a different expression from the all-to-all coupled case. 
{Their explicit form is reported in Appendix \ref{Appendix: Order-by-disorder}.
\begin{widetext}
At small reciprocal regime $|j_+| \ll |j_-|,j_0$, 
one finds,
\begin{eqnarray}
\label{eq: Delta phi dynamics spatial}
 \Delta\dot\phi&\simeq&\bigg[-2j_+
 + \frac{ j_-^2\sigma^2
 \cos\Delta\phi}{4j_0}
 \sum_{\bm k,\bm {k'}}
 \frac{j_0^2{\bm k'}^2
 -2 j_-^2\cos^2\Delta\phi}
 {(4j_-^2\cos^2\Delta\phi-j_0^2\bm k^2)(4j_-^2\cos^2\Delta\phi-j_0^2{\bm k'}^2)\bm k^2 {\bm k'}^2}
 \bigg] \sin\Delta\phi,
\end{eqnarray}
which has a similar form to the all-to-all coupling case (Eq.~\eqref{eq: phase order dynamics renormalized small j+}).
Especially when $j_+=0$ (perfectly non-reciprocal limit), one can readily check that Eq.~\eqref{eq: Delta phi dynamics spatial} has a stable fixed point at $\Delta\phi_*=\pm\pi/2$ corresponding to the chiral phase (as in the previous section).
Therefore, the second term describes the ``entropic torque'' that drives the system to the spontaneous parity-broken phase, which competes with the reciprocal coupling (first term of Eq.~\eqref{eq: Delta phi dynamics spatial}) that drives the system to the static, parity-symmetric phase.
This triggers a non-reciprocal phase transition \cite{Fruchart2021} from a static phase $\Delta\phi_*=0,\pi$ to the chiral phase $\Delta\phi_*\ne 0,\pi$ as the noise strength is \textit{increased}, again, signaling the OBDP.
\end{widetext}




We briefly note that the fluctuations is known to exhibit an anomalous enhancement that diverges for spatial dimensions below $d=4$ at the phase transition point of a non-reciprocal phase transition \cite{Hanai2020}.
A recent study \cite{Zelle2023} has further demonstrated that these significant fluctuations induce a discontinuous transition for spatial dimensions below $d=4$. To fully capture this physics, a more advanced analysis beyond the lowest-order perturbative approach employed in this study is necessary, which is beyond the scope of this paper.
We emphasize, however, that our noise-induced spontaneous symmetry-breaking scenario itself would be unaffected by this fluctuation physics, as the OBDP physics is relevant outside the critical regime.

\begin{figure*}
\centering
\includegraphics[width=0.96\linewidth,keepaspectratio]{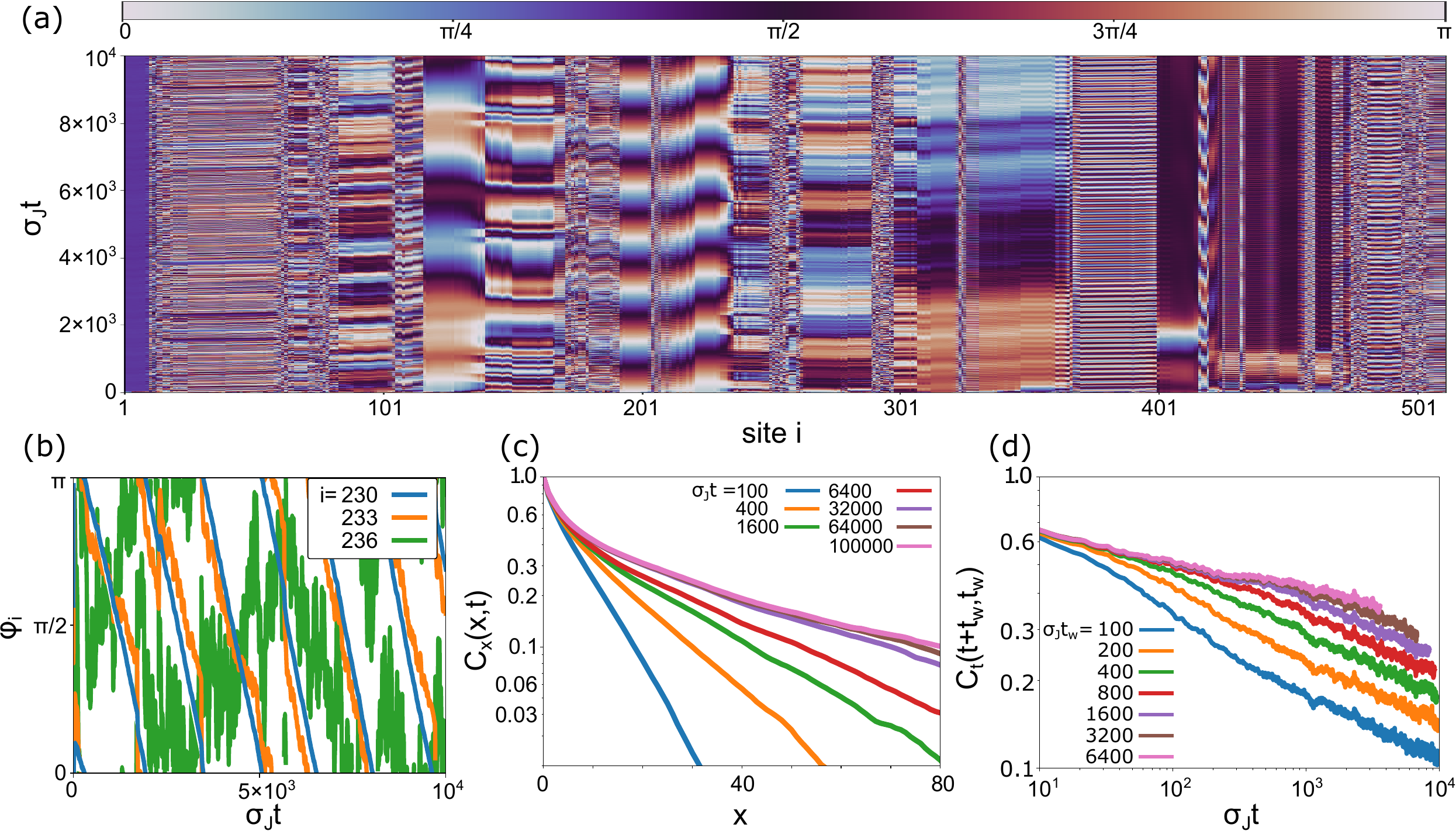}
\caption{
\textbf{Non-reciprocal frustration induced spin-glass-like state in an asymmetric random spin chain.}
The asymmetric coupling case ($J_{ij}\ne J_{ji}$) with the coupling to the left $J_i^{\rm L}$ and the right $J_i^{\rm R}$ sampled independently ($\gamma=0$).
(a),(b) Typical trajectory of (nematic) angles $\varphi_i=\theta_i({\rm mod}~\pi)$ of a one-dimensional random non-reciprocal spin chain. 
Here, we have set $N=2^9=512$ and the initial condition was taken randomly from a uniform distribution $\theta_i=[0,2\pi)$.
(b) Line-cut data of the trajectory at site $i=230,233,236$ of (a). 
(c) Spatial correlation function $C_x(x,t)$. 
(d) Time correlation function $C_t(t_w+t,t_w)$. Note that both axes are plotted on a logarithmic scale. 
We have averaged over $500$ trajectories for $\sigma_J t\le 6.4\times 10^3$ in panels (c) and $400$ trajectories for $\sigma_J t\ge 3.2\times 10^4$ in (c) and for all plots in (d).
We have set $N=2^{10}=1024$. 
This system exhibits slow dynamics characterized by power-law decay and aging phenomena and a short-ranged spatial correlation (which exhibits stretched exponential decay, see Fig.~\ref{fig: stretched exponential}), where the latter property is in stark contrast to the nematically ordered state seen in the reciprocal case (Fig.~\ref{fig: glass reciprocal}).
These properties are reminiscent of a spin glass.}
\label{fig: glass_asymmetric}
\end{figure*}

\section{Non-reciprocity induced spin-glass-like state}
\label{sec:spin glass}


Another striking phenomenon arising from geometrical frustration is the emergence of spin glasses \cite{Edwards1975,Sherrington1975,Parisi1979,Sompolinsky1982,Cugliandolo1993,Fisher1998a,Fisher1998b}, which occurs ubiquitously in geometrically frustrated systems with random interactions. 
In such a situation, a macroscopic number of fixed points and saddle points are generated to make the energy landscape bumpy. This makes it extremely difficult for the system to find its global minimum, resulting in slow dynamics characterized by a power law decay (or slower \cite{Fisher1998b}) of time correlation functions and the aging phenomena \cite{Sompolinsky1982,Cugliandolo1993,Fisher1998b} associated with no long-ranged spatial order.

A natural question is whether such glassy states can be generated by non-reciprocal interaction. 
It is tempting to expect the negative, as they induce the chase-and-runaway dynamics that may cause the glass to melt.
Indeed, there are a number of works that support this view \cite{Crisanti1988,Rieger1988,Rieger1989,Parisi1986,Hertz1987,Sompolinsky1986,Sompolinsky1988,Rieger_eco1989,Bunin2017} including the works in the context of neural \cite{Rieger1988,Parisi1986,Hertz1987,Sompolinsky1986,Sompolinsky1988} and ecological systems \cite{Rieger_eco1989,Bunin2017}.
However, the above studies analyzed (mostly all-to-all coupled) models that already contained geometrical frustration in the reciprocal limit, making the exact role of non-reciprocal frustration unclear. 

To unambiguously study the effect of non-reciprocal frustration alone, it is important for us to consider models that have \textit{no} geometrical frustration in the reciprocal limit.
For this purpose, we consider a one-dimensional XY spin chain that follows Eq.~\eqref{eq: XY-model} that consists of $N$ spins with nearest-neighbor interaction
$J_{ij} = J_{i}^{\rm R}\delta_{i+1,j}+J_{i}^{\rm L}\delta_{i,j+1}$ in an open boundary condition (Fig.~\ref{fig: glass model}),
\begin{eqnarray}
    \label{eq: 1D asymmetric random chain}
    \dot\theta_i 
    = J_i^{\rm R} \sin(\theta_{i+1}-\theta_i)
    +J_i^{\rm L}
    \sin(\theta_{i-1}-\theta_i),
\end{eqnarray}
with $J_{i}^{\rm L/R}$ being randomly distributed according to 
\begin{eqnarray}
\label{eq: random spin chain distribution}
    p(J_i^{\rm L/R})\propto 
    \begin{cases}
        e^{-(J_i^{\rm L/R})^2/(2\sigma_J^2)}
    & |J_i^{\rm L/R}| \ge  J_c \\
    0 & |J_i^{\rm L/R}| < J_c
    \end{cases}
\end{eqnarray}
Here, $\sigma_J^2= \avg{(J_{i
}^{\rm R})^2}=\avg{(J_{i
}^{\rm L})^2}$ characterizes the randomness of the coupling and has introduced a cutoff $J_c$ (which we set $J_c=0.1\sigma_J$ throughout)
to prevent the coupling from completely vanishing. 
We also introduce an asymmetry parameter $\gamma$ defined by
$\avg{J_{i}^{\rm L}J_{i}^{\rm R}}\equiv\gamma\sigma_J^2$
that parameterize the asymmetry (non-reciprocity) of the coupling.
For example, $\gamma=1$ corresponds to the reciprocal limit (where all spins satisfy $J_{ij}=J_{ji}$),
while $\gamma=-1$ corresponds to the anti-symmetric limit (where all spins satisfy $J_{ij}=-J_{ji}$).
$\gamma=0$ corresponds to the case where $J_{ij}$ and $J_{ji}$ are independent.


This model has a crucial advantage in that no geometrical frustration exists
in the reciprocal case $\gamma=1$, 
and therefore, frustration can only arise through non-reciprocal interactions. 
This can be seen from the fact that the ground state configuration of the reciprocal system is uniquely determined once one fixes one of the spins (Fig.~\ref{fig: glass reciprocal} bottom-left panel).
Since reciprocal coupling favors either alignment or anti-alignment of spins, the ground state in the reciprocal limit exhibits a nematic order characterized by a complex order parameter $\psi_2 = (1/N)\sum_{i=1}^N e^{2i\theta_i}$.

Figure \ref{fig: glass reciprocal}(a) shows a typical trajectory of $\varphi_i=\theta_i({\rm mod}~\pi)$ [which regards the angles of the arrow pointing to opposite directions as being identical, thus making it useful to measure nematicity] in the reciprocal case, $J_i^{\rm R}=J_i^{\rm L}$.
Here, we have set the initial state to be random.
As seen, the dynamics are governed by the annihilation dynamics of the initially created (nematic) domain walls (Figs.~\ref{fig: glass reciprocal}(b)) towards the nematic long-range ordered state. 
This is captured in the spatial correlation function 
\begin{eqnarray}
\label{eq: spatial correlation function}
    C_{x}(x,t)=\left|\overline{\frac{1}{N-x}\sum_{i=1}^{N-x}
    \psi_{2,i+x}(t)\psi_{2,i}^*(t)}\right|    
\end{eqnarray}
that is converging towards the long-ranged ordered state $C_x(x,t\rightarrow\infty)\rightarrow 1$ (Fig.~\ref{fig: glass reciprocal}(b)).
Here, $\psi_{2,i}(t)=e^{2i\theta_i(t)}$ is a complex representation of nematic direction at site $i$, and $\overline{(\cdots)}$ represents the average over random initial conditions, with a different configuration of $J_{ij}$ taken for each run.  
The temporal correlation function
\begin{eqnarray}
    \label{eq: temporal correlation function}
    C_t(t_w+t,t_w)=\left|\overline{\frac{1}{N}\sum_{i=1}^N \psi_{2,i}(t_w+t)\psi_{2,i}^*(t_w)}\right|
\end{eqnarray}
in this phase is expected to converge to a constant. 
Note that, because of the slow domain-wall annihilation process, the temporal correlation function $C(t+t_w,t_w)$
exhibits an aging behavior, i.e., the feature that the system takes more time to decorrelate as the waiting time $t_w(\gg \sigma_J^{-1}, J_c^{-1})$ proceeds (Fig.~\ref{fig: glass reciprocal}(c)).




Now let us turn to the non-reciprocal case $\gamma=0$,
where we sample the couplings $J_i^{\rm R}$ and $J_i^{\rm L}$  independently.
In this case, as seen in Fig.~\ref{fig: glass_asymmetric}(a), we observe the formation of domains that are \textit{locally} nematically ordered, in which many of them are almost time periodic (see e.g., $i=230$ in Fig. \ref{fig: glass_asymmetric}(b)) but others seem to be interrupted ($i=233$) by the nearby chaotic domain ($i=236$).
These behaviors are vastly different from the reciprocal case of Fig.~\ref{fig: glass reciprocal}(a) dominated by domain wall annihilation dynamics. 


\begin{figure}
\centering
\includegraphics[width=0.75\linewidth,keepaspectratio]{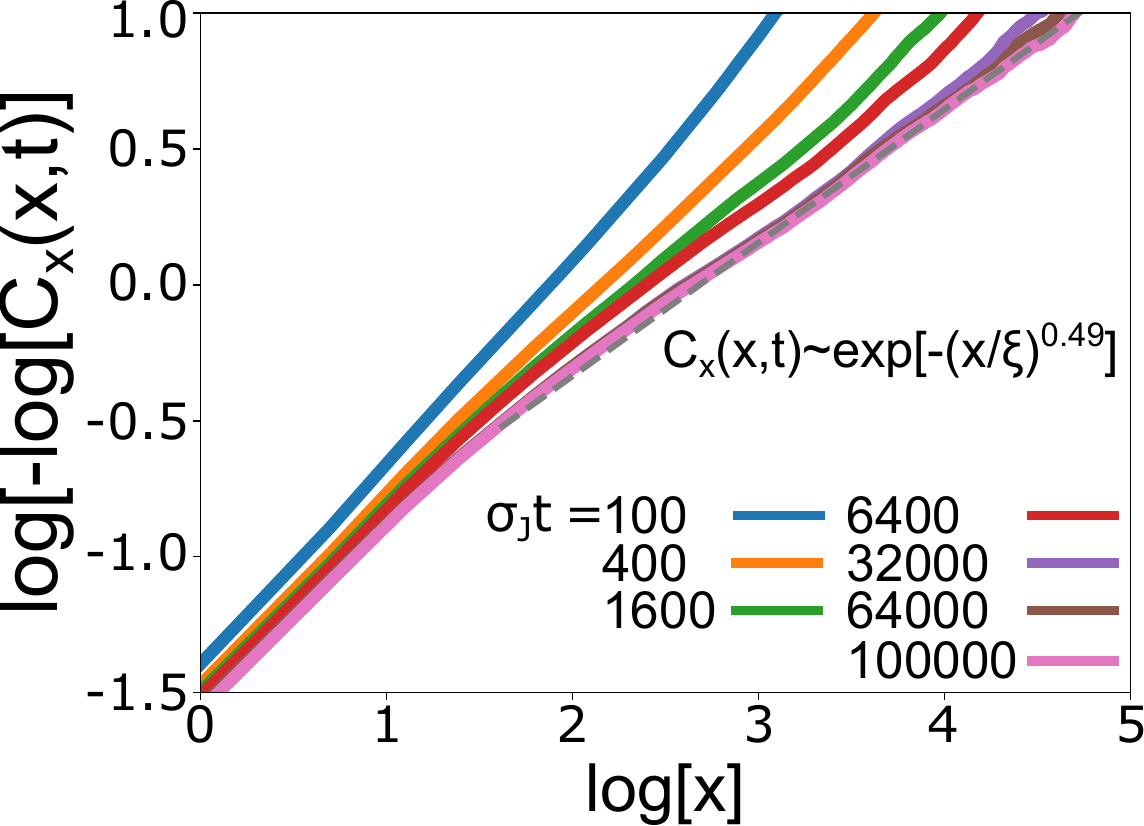}
\caption{
\textbf{Stretched exponential decay of spatial correlation function in an asymmetric random spin chain.}
This is identical to Fig.~\ref{fig: glass_asymmetric}(c) but plotted in a different scale.
The dashed line is a fitted curve $C_x(x,t)=e^{-(x/\xi)^\alpha}$ to the $\sigma_J t=10^5$ result, which gives $\alpha\simeq 0.49$.
}
\label{fig: stretched exponential}
\end{figure}

\begin{figure*}
\centering
\includegraphics[width=0.95\linewidth,keepaspectratio]{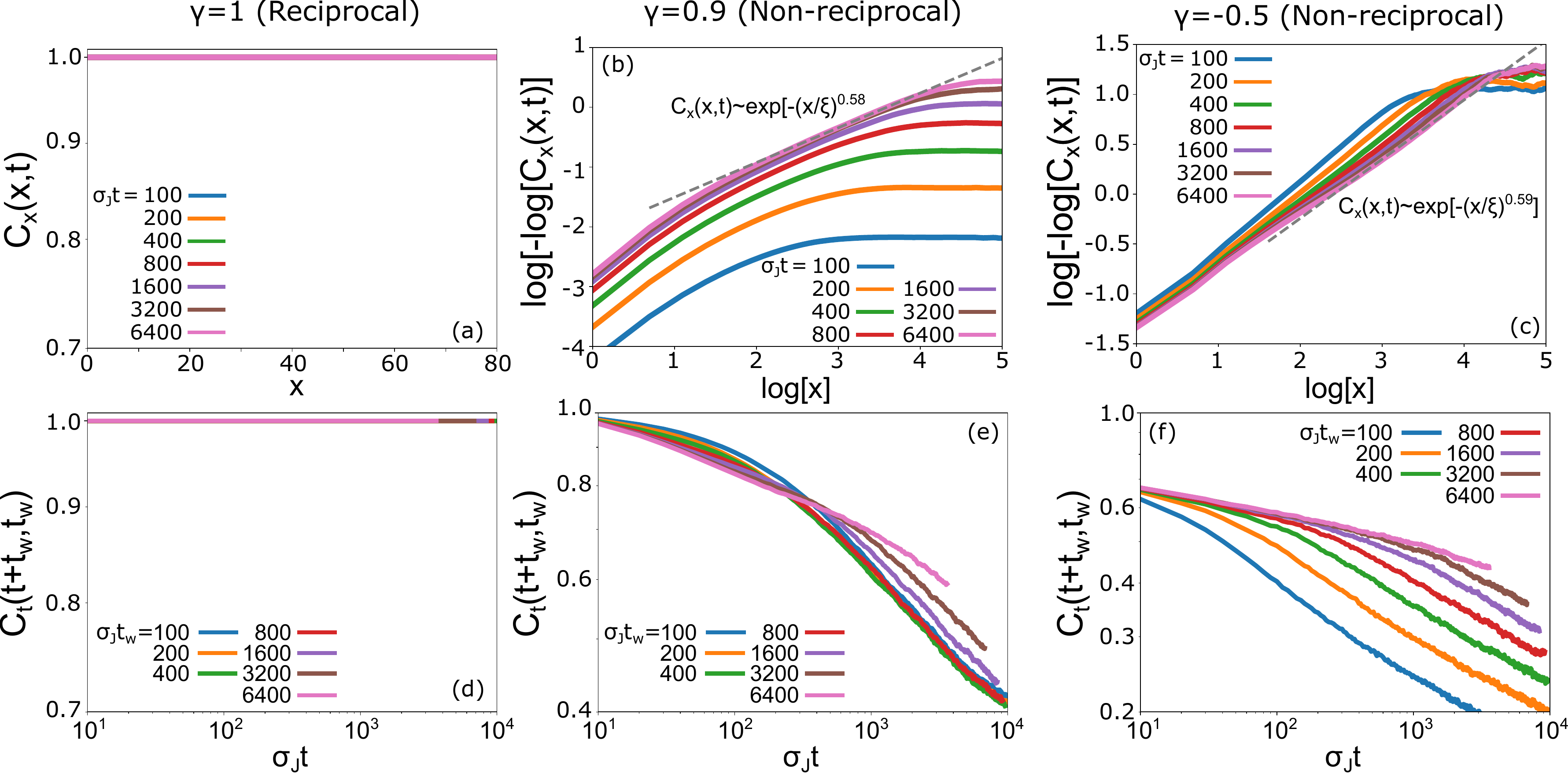}
\caption{
\textbf{Asymmetry parameter $\gamma$ dependence of the spatial and time correlation functions.}
(a)-(c) Spatial correlation function $C_x(x,t)$ and (d)-(f) temporal correlation function $C_t(t+t_w,t_w)$.
(a),(d) $\gamma = 1$. (b), (e) $\gamma=0.9.$ (c), (f) $\gamma = -0.5$. 
We set $N=500$ in panels (a),(d) and $N=1000$ in panels (b),(c),(e),(f) and averaged over 500 trajectories. 
The initial state is set to be close to a nematic phase $\theta_i = \sum_{j=1}^i \pi(1+{\rm sgn}(J_j^{\rm R}))/2+10^{-3}\eta_i$, where $\eta_i=[0,2\pi)$ is a uniformly distributed random valuable.}
\label{fig: glass gamma dependence}
\end{figure*}

\begin{figure}
\centering
\includegraphics[width=0.75\linewidth,keepaspectratio]{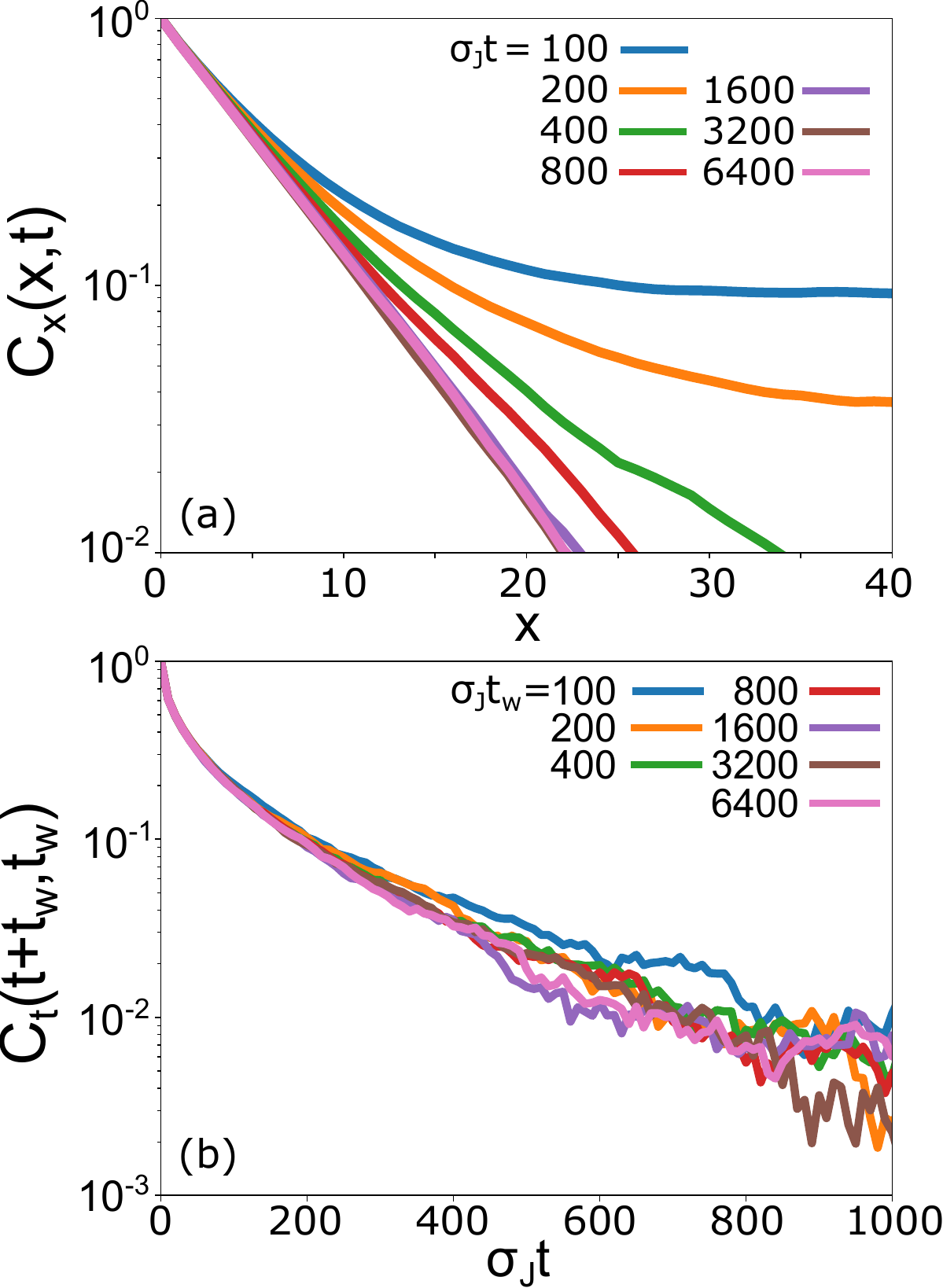}
\caption{
\textbf{Correlation function in the disordered state.}
(a) Spatial correlation function $C_x(x,t)$ and (b) temporal correlation function $C_t(t+t_w,t_w)$.
We set $\gamma=1$ and the noise strength $\sigma=0.1$. We have averaged over 500 trajectories.
Similarly to Fig.~\ref{fig: glass gamma dependence}, we have set the initial state to be close to a nematic phase $\theta_i = \sum_{j=1}^i \pi(1+{\rm sgn}(J_j^{\rm R}))/2+10^{-3}\eta_i$, where $\eta_i=[0,2\pi)$ is a uniformly distributed random valuable.}
\label{fig: disordered state correlation function}
\end{figure}

Figures \ref{fig: glass_asymmetric}(c) and (d) show the spatial and time correlation function of this asymmetric spin chain, respectively.
Strikingly, the time correlation function exhibits a power-law decay $C_t(t_w+t,t_w)\sim t^{-\alpha}$ at large $t$ with a clear sign of aging, while the state is converging towards a short-ranged spatially correlated state (which exhibits a stretched exponential decay $C_x(x,t)\sim e^{-(|x|/\xi)^{\alpha}}$ with $\alpha=0.49$, see Fig.~\ref{fig: stretched exponential}), in stark contrast to the reciprocal case.
These features are reminiscent of a spin glass, except that the time correlation function does not seem to converge to a finite value at $t\rightarrow\infty$ \cite{Edwards1975} (at least up to $\sigma_J t = 10^5$), 
implying that the state does not completely freeze to a static state.


Figure~\ref{fig: glass gamma dependence} shows the spatial and temporal correlation function for various values of asymmetry parameter $\gamma$.
Here, to make the convergence to the nematic state in the nematic order phase faster, we have chosen the initial states to be close to the nematic phase. 
We observe that even at very weak non-reciprocity ($\gamma=0.9$), qualitatively the same feature to the spin-glass-like phase observed at $\gamma=0$ in Fig.~\ref{fig: glass_asymmetric} is also seen.

This slow decay observed in asymmetric cases is qualitatively different from the disordered state that occurs when stochastic noise $\eta_i(t)$ is added to Eq.~\eqref{eq: 1D asymmetric random chain}.
Figure~\ref{fig: disordered state correlation function} shows the spatial and temporal correlation functions in the presence of a small Gaussian white noise (where $\avg{\eta_i(t)}=0,\avg{\eta_i(t)\eta_j(t')}=\sigma \delta_{ij}\delta(t-t')$).
As seen, both the spatial and temporal correlation function at the long time limit exhibit an exponential decay in the disordered phase, in qualitative difference from our spin-glass-like state in Fig.~\ref{fig: glass_asymmetric}.

Summarizing, we have found numerical evidence that non-reciprocal coupling can induce a phase reminiscent of a spin-glass, which is qualitatively different from both the disordered state and the nematically ordered state.
For convenience, we have summarized the different behaviors of the correlation function in different phases in Table \ref{table: phases of asymmetric spin chain}.
In contrast to both of these phases, the spin-glass-like state exhibits a algebraic decay in the temporal correlation function while exhibiting a short spatial correlation.

\begin{table*}
    \caption{Phases in an asymmetric random spin chain and the behaviors of correlation function}
    \label{table: phases of asymmetric spin chain}
    \centering
    \begin{tabular}{c|c|c|c}
        Parameter &
        Phase
        & Spatial correlation function &
        Temporal correlation function
        \\
        \hline
        Reciprocal ($\gamma=1$)
        	\& deterministic
        & Nematically ordered
        &
        Long-ranged order
        &
        Converges to a constant (slowly)
        \\
        Non-reciprocal	($\gamma<1$)
        \& deterministic
        & Spin-glass-like
        &
        Stretched exponential decay
        &
        Power law decay with aging
        \\
        Stochastic
        & Disordered
        &
        Exponential decay
        &
        Exponential decay without aging
        
    \end{tabular}
\end{table*}

We remark that a similar slow decay to the one observed in our spin-glass-like phase has been observed in one-dimensional coupled logistic maps in their discrete-time evolution, as pioneered by Kaneko \cite{Kaneko1989,Kaneko1992}.
In his model, each site is itself a logistic map that exhibits bifurcations to limit cycles or chaos, 
and these sites are coupled with their neighboring sites.  
At a phenomenological level, we observe interesting similarities between our model and Kaneko's model: 
in the former, by regarding each domain seen in Fig.~\ref{fig: glass_asymmetric}(a) as a chaotic or periodic element, each element seems to be attempting to align with the nearby domains, somewhat analogous to the latter situation.
However, there are also clear differences, e.g., the randomness is explicitly encoded in the former from random coupling (similarly to the original spin glass problem) while they are generated spontaneously from chaos in the latter. 
The connection between the two models deserves further investigation.

\section{Conclusion and outlook}
\label{sec:conclusion}



We have shown that non-reciprocal interaction may generate marginal orbits (``accidental degeneracy'') similar to those in geometrically frustrated systems, establishing a direct analogy between the two classes of systems. 
We have shown that the emergence of this ``accidental degeneracy'' can give rise to a dynamical counterpart of order-by-disorder phenomena and spin glasses. 
Our results offer an unexpected bridge between complex magnetic materials with geometrical frustration and non-reciprocal systems. 

There are many possible directions for further extensions.
For example, in Sec.~\ref{sec:order-by-disorder}, where we studied order-by-disorder phenomena induced by non-reciprocal frustration, for simplicity, we have focused on models that have accidental degeneracy of orbits parameterized by just one or two parameters in the deterministic limit. 
This is in a similar situation to (geometrically frustrated) 2D $J_1$-$J_2$ XY model \cite{Henley1989}.
However, there are various geometrically frustrated models that have \textit{macroscopic} a number of ground-state degeneracy. They either exhibit order-by-disorder \cite{Villain1980, Moessner1998a, Moessner1998b} or classical spin liquids that lack long-ranged order in the low-temperature limit $T\rightarrow 0$, depending on the number of unconstrained degrees of freedom and number of directions of gapless excitations~\cite{Moessner1998a, Moessner1998b}. 
We plan to study systems with such macroscopic numbers of degeneracy induced by non-reciprocal frustrations \cite{Wiesenfeld1989} to find the criterion for the emergence of order-by-disorder similar to those given in Refs.~\cite{Moessner1998a, Moessner1998b} for geometrically frustrated systems.

For the spin-glass-like state that was studied in Sec.~\ref{sec:spin glass}, the purpose of focusing on a one-dimensional spin chain with nearest-neighbor interactions was to make sure that geometrical frustration is absent and therefore the glassy dynamics observed in the simulation can be safely attributed to non-reciprocal effects.
In two or higher dimensions, the two types of frustrations (geometrical and non-reciprocal) may co-exist.
It would be interesting to ask how the conventional spin-glass (that is characterized by a non-zero Edwards-Anderson order parameter \cite{Edwards1975}, i.e., $q_{\rm ED}=\lim_{t\rightarrow\infty}\lim_{t_w\rightarrow\infty} C_t(t_w+t,t)\ne 0$) evolves to the non-reciprocal spin-glass like state (that exhibits a vanishing $q_{\rm ED}=0$ but still exhibits aging phenomena found in this work, and what are the properties of the phase transitions between them, if any.

Another possible direction is to extend our work to open quantum systems~\cite{Metelmann2015, Lee2014}. 
In Ref.~\cite{Metelmann2015}, a recipe to realize non-reciprocal interaction via reservoir engineering has been proposed and non-reciprocal hopping has already been implemented experimentally~\cite{Fang2017}. 
It would be interesting to ask whether states reminiscent of quantum spin liquids with long-ranged entanglement can appear by non-reciprocal frustration.

Finally, in this work, we have focused on how far we can push the \textit{analogies} between geometrical and non-reciprocal frustrations. Given that we have established such analogies, an interesting next step would be to ask what the fundamental \textit{differences} between the two is, other than the rather obvious difference that the final states are usually time-dependent in the latter.

\textbf{Acknowledgement.}
RH thanks Mark Bowick, Fridtjof Brauns, Aashish Clerk, Michel Fruchart, Ramin Golestanian, Hisao Hayakawa, Peter B. Littlewood, M. Cristina Marchetti, Alexander McDonald, Roderich Moessner, Vincenzo Vitelli, Cheyne Weis, and Fr\'ed\'eric van Wijland for discussions. 
He also thanks Samuel Begg for the critical reading of the manuscript.
This work was supported by an appointment to the JRG Program at the APCTP through the Science and Technology Promotion Fund and Lottery Fund of the Korean Government and by Grant-in-Aid for 	
Grant-in-Aid for Research Activity Start-up from
JSPS in Japan (No. 23K19034).










\appendix

\section{Liouville-type theorem for perfectly non-reciprocal systems}
\label{Appendix: Liouville-type theorem}

\subsection{XY-model}

Here we provide the proof for the Liouville-type theorem (Eq.~\eqref{eq: Liouville's theorem} in the main text) for the XY-model
\begin{eqnarray}
    \label{SIeq: XY-model}
    \dot\theta_i = -\sum_j J_{ij}\sin(\theta_i-\theta_j).
\end{eqnarray}
We also provide its generalization to more general non-reciprocal models.
The continuity equation of the distribution function $\rho$ for the XY-model (Eq.~\eqref{SIeq: XY-model}) is given by
\begin{eqnarray}
    \label{SIeq: continuity equation}
    \frac{\partial\rho}{\partial t}
    =-\sum_i \frac{\partial(\rho\dot\theta_i)}{\partial\theta_i}
    =-\sum_{i}\Big[
    \frac{\partial\rho}{\partial\theta_i}\dot\theta_i
    +\rho\frac{\partial\dot\theta_i}{\partial\theta_i}
    \Big].
\end{eqnarray}
In the perfectly non-reciprocal case $J_{ij}=-J_{ji}$, the second term of Eq.~\eqref{SIeq: continuity equation} can be shown to vanish as,
\begin{eqnarray}
    \rho\sum_i
    \frac{\partial\dot\theta_i}{\partial\theta_i}
    =\rho\sum_{ij}\big[
    J_{ij}\cos(\theta_i-\theta_j)
    \big] = 0, 
\end{eqnarray}
where in the last equality, we have used the property that $J_{ij}$ is anti-symmetric and $\cos(\theta_i-\theta_j)$ is symmetric.
This gives 
\begin{eqnarray}
    \frac{d\rho}{dt}
    = \frac{\partial\rho}{\partial t}
    +\sum_i \frac{\partial\rho}{\partial\theta_i}\dot\theta_i=0,
\end{eqnarray}
proving the Liouville-type theorem.

\subsection{Heisenberg model}

In the above proof, note how we have only used the property that the derivative of the right-hand side of the dynamical system Eq.~\eqref{SIeq: XY-model} is anti-symmetric.
This suggests that the Liouville-type theorem holds more generally.
For example, the Heisenberg spin $\bm S_i=(S_i^x,S_i^y,S_i^z)$ (with $|\bm S_i|^2=1$) systems that is described by the Landau-Lifshitz equation \cite{Aharoni1996},
\begin{eqnarray}
\label{Method eq: Heisenberg}
    \dot {\bm S}_i = -\sum_{j=1}^N J_{ij} 
    [\bm S_i \times \bm S_j
    +  \alpha \bm S_i \times (\bm S_i\times \bm S_j)]
\end{eqnarray}
can be shown to satisfy the Liouville-type theorem 
\begin{eqnarray}
\label{Method eq: Liouville Heisenberg}
    \frac{d\rho}{dt}
    = \frac{\partial\rho}{\partial t}
    +\sum_{i=1}^N\sum_{\mu=x,y,z}
    \frac{\partial \rho}{\partial S_i^\mu}
     \dot S_i^\mu
\end{eqnarray}
when the coupling is anti-symmetric  $J_{ij}=-J_{ji}$.
This can be shown by noting that
\begin{eqnarray}
\label{Method eq: Current conservation Heisenberg}
    \frac{\partial\rho}{\partial t}
    = -\sum_{i,\mu}
    \frac{\partial (\rho \dot S_i^\mu) }{\partial S_i^\mu} 
    = -\sum_{i,\mu}\Big[
    \frac{\partial \rho}{\partial S_i^\mu}
    \dot S_i^\mu
    + \rho \frac{\partial \dot S_i^\mu}{\partial S_i^\mu} 
    \Big].
\end{eqnarray}
Rewriting Eq.~\eqref{Method eq: Heisenberg} as (where $\epsilon_{\mu\nu\sigma}$ is the Levi-Civita symbol)
\begin{eqnarray}
    \dot S_i^\mu 
    = -\sum_j J_{ij}
    \Big[
    \sum_{\nu\sigma}
    \epsilon_{\mu\nu\sigma}S_i^\nu S_j^\sigma
    +\alpha \big[\sum_\nu S_i^\nu S_j^\nu S_i^\mu - S_j^\mu\big]
    \Big],
    \nonumber\\
\end{eqnarray}
one can show that
\begin{eqnarray}
\label{Method eq:  d dot S / dS}
    \sum_{i,\mu}\frac{\partial \dot S_i^\mu}{\partial S_i^\mu}
    &= &
    \sum_{i,j,\mu} J_{ij}
    \frac{\partial}{\partial S_i^\mu }
    \Big[
    \sum_{\nu\sigma}
    \epsilon_{\mu\nu\sigma}S_i^\nu S_j^\sigma
    \nonumber\\
    && \ \ \ \ \ \ \ \ \ \ \ 
    +\alpha \big[\sum_\nu S_i^\nu S_j^\nu S_i^\mu - S_j^\mu\big]
    \Big]
    \nonumber\\
    &=&
    \sum_{i,j,\mu} J_{ij}
    \Big[
    \sum_{\nu\sigma}
    \epsilon_{\mu\nu\sigma}\delta_{\mu\nu} S_j^\sigma
    \nonumber\\
    && \ \ \ \ \ \ \ \ \ \ \ 
    +\alpha \sum_\nu \big[\delta_{\mu\nu} S_j^\nu S_i^\mu 
    + S_i^\nu S_j^\nu \big]
    \Big]    \nonumber\\
    &=&
    \sum_{i,j,\mu} J_{ij}\alpha \big[S_i^\mu S_j^\mu 
    + \sum_\nu  S_i^\nu S_j^\nu \big] 
    \nonumber\\
    &=&0,
\end{eqnarray}
holds, where again, we have used the property that $J_{ij}$ is anti-symmetric $J_{ij}=-J_{ji}$ in the last line.
Combining Eqs.~\eqref{Method eq: Current conservation Heisenberg} and \eqref{Method eq:  d dot S / dS} proves the Liouville-type theorem for perfectly non-reciprocal Heisenberg system (Eq.~\eqref{Method eq: Liouville Heisenberg}).

\subsection{Coupled oscillators with phase delay}

We consider a system composed of coupled oscillators with a phase delay \cite{Sakaguchi1986, Ernest2011}, \begin{eqnarray}
    \dot\theta_i 
    = 
    \omega_i+\sum_j J_{ij} \sin(\theta_j - \theta_i + \alpha_i).
\end{eqnarray}
Here, $0\le \alpha\le \pi/2$ is the phase delay, $\omega_i$ is a natural frequency, and the coupling constant $J_{ij}=J_{ji}$ is symmetric.
This model is relevant for 
the physics of biased Josephson junctions arrays \cite{Wiesenfeld1996, Wiesenfeld1998} and microscopic rotors \cite{Uchida2010} that can be derived from microscopic models.

The phase delay $\alpha_i\ne 0$ drives the coupling to be non-reciprocal.
In the non-reciprocal limit $\alpha_i=\pi/2$, 
\begin{eqnarray}
    \dot\theta_i 
    =\omega_i -\sum_j J_{ij} \cos(\theta_i - \theta_j),
\end{eqnarray}
the Liouville-type theorem
\begin{eqnarray}
    \label{eq: Liouville's theorem Sakaguchi-Kuramoto}
    \frac{d\rho}{dt}
    = \frac{\partial\rho}{\partial t}
    +\sum_i \frac{\partial\rho}{\partial\theta_i}\dot\theta_i=0
\end{eqnarray}
holds.
This can be shown from the relation
\begin{eqnarray}
    \rho\sum_i
    \frac{\partial\dot\theta_i}{\partial\theta_i}
    =\rho J
    \sum_{i,j}
    \sin(\theta_i-\theta_j)
    =0
\end{eqnarray}
that gives
\begin{eqnarray}
    \frac{\partial\rho}{\partial t}
    =-\sum_i \frac{\partial(\rho\dot\theta_i)}{\partial\theta_i}
    =-\sum_i \frac{\partial\rho}{\partial\theta_i}
    \dot\theta_i
\end{eqnarray}
and hence Eq.~\eqref{eq: Liouville's theorem Sakaguchi-Kuramoto} is proven.

\subsection{Non-reciprocally interacting particles}

Another example is the non-reciprocally interacting particles that are realized in systems such as complex plasma \cite{Ivlev2015} and chemically \cite{Soto2014, Saha2019} and optically active colloidal matter \cite{Yifat2018, Parker2020}.
The position of the particle $\bm r_i=(x_i,y_i,z_i)$ of an interacting system is given by, 
\begin{eqnarray}
    \dot {\bm r}_i =  \sum_{j} \bm f_{ij}(|\bm r_i - \bm r_j|),
\end{eqnarray}
where the force $\bm f_{ij}(|\bm r_i-\bm r_j|)$ acting on the particle $i$ from the interaction with the particle $j$ is assumed to be a function of the inter-particle distance $|\bm r_i-\bm r_j|$.
The force can in general split into reciprocal and anti-reciprocal contributions,
\begin{eqnarray}
    \bm f_{ij}=\bm f_{ij}^r(|\bm r_i - \bm r_j|)
    +\bm f_{ij}^a(|\bm r_i - \bm r_j|)
\end{eqnarray}
where $\bm f_{ij}^r(|\bm r_i - \bm r_j|)=-\bm f_{ji}^r(|\bm r_i - \bm r_j|)$ and $\bm f_{ij}^a(|\bm r_i - \bm r_j|)=\bm f_{ji}^a(|\bm r_i - \bm r_j|)$ are the reciprocal and anti-reciprocal contributions, respectively.

In the anti-reciprocal case 
$\bm f_{ij}=\bm f_{ij}^a(|\bm r_i - \bm r_j|)$, Liouville-type theorem holds.
Similar to the spin systems,
the continuity equation reads
(where $\nabla_i=(\partial_{x_i},\partial_{y_i},\partial_{z_i})$)
\begin{eqnarray}
    \label{SIeq: continuity equation non-reciprocal particles}
    \frac{\partial\rho}{\partial t}
    =-\sum_i  \nabla_i\cdot(\rho\dot{\bm r}_i)
    =-\sum_{i} \Big[
    (\nabla_i\rho)\cdot \dot{\bm r}_i
    +\rho \nabla_i \cdot \dot{\bm r}_i
    \Big].
    \nonumber\\
\end{eqnarray}
The last term of Eq.~\eqref{SIeq: continuity equation non-reciprocal particles} can be shown to vanish, 
\begin{eqnarray}
    \rho\sum_i
    \nabla_i\cdot\dot{\bm r}_i
    =\rho\sum_{ij}\big[
    \nabla_i \cdot\bm f_{ij}^a(|\bm r_i-\bm r_j|
    \big] = 0, 
\end{eqnarray}
since
\begin{eqnarray}
\sum_{ij}\nabla_i\cdot\bm f_{ij}^a(|\bm r_i-\bm r_j| 
&=& \sum_{ij}\nabla_j \cdot\bm f_{ji}^a(|\bm r_j-
\bm r_i|)\nonumber\\
&=& \sum_{ij}
\nabla_j \cdot\bm f_{ij}^a(|\bm r_i-\bm r_j|)\nonumber\\
&=& -\sum_{ij}
\nabla_i \cdot\bm f_{ij}^a(|\bm r_i-\bm r_j|).
\nonumber\\
\end{eqnarray}
This proves the desired Liouville-type theorem 
\begin{eqnarray}
    \label{SIeq: Liouville theorem non-reciprocal particles}
    \frac{\partial\rho}{\partial t}
    +\sum_{i} 
    (\nabla_i\rho)\cdot \dot{\bm r}_i = 0,
\end{eqnarray}
for non-reciprocally interacting systems with anti-symmetric coupling.





\section{Order-by-disorder phenomena} 
\label{Appendix: Order-by-disorder}

\subsection{All-to-all coupled model}
We provide here the details of the analysis of order-by-disorder phenomena (OBDP) occurring in both geometrically and non-reciprocally frustrated systems.
For concreteness, we consider the dynamics of all-to-all coupled XY-model grouped into a few communities $a={\rm A,B,C,...},$ following the Langevin equation,
\begin{eqnarray}
    \label{SIeq: Langevin order-by-disorder}
    \dot\theta_i^a =
    - \sum_b
    \frac{j_{ab}}{N_b}
    \sum_{j=1}^{N_b} 
    \sin (\theta_i^a - \theta_j^b)
    + \eta_i^a,
\end{eqnarray}
where $\avg{\eta_i^a(t)}=0,\avg{\eta_i^a(t)\eta_j^b(t')}=\sigma\delta_{ab}\delta_{ij}\delta(t-t')$.
The all-to-all coupled nature allows us to rewrite Eq.~\eqref{SIeq: Langevin order-by-disorder} in a single spin picture,  
\begin{eqnarray}
    \dot\theta_i^a = -\sum_{b}j_{ab}r_b\sin(\theta_i^a-\phi_b)
    +\eta_i^a,
\end{eqnarray}
by introducing the order parameter $\psi_a = (1/N_a)\sum_i^{N_a}e^{i\theta_i^a}=r_a e^{i\phi_a}$. 

As emphasized in the main text, when the inter-community coupling is taken to be geometrically/non-reciprocally frustrated, the order parameter dynamics can take different orbits $\phi(t)=(\phi_{\rm A}(t),\phi_{\rm B}(t),...)$ depending on their initial condition in the absence of stochasticity.
We will show below that this ``accidental degeneracy'' of orbits is generically lifted by the presence of noise. 

To proceed, we consider the dynamics of fluctuations $\delta\theta_i^a = \theta_i^a - \phi_a$ caused by noise. Assuming weak noise strength, we linearize the stochastic equation of motion as
\begin{eqnarray}
\label{SIeq: fluctuation dynamics}
    \delta\dot\theta_i^a
    \approx -\sum_b j_{ab}
    \cos(\phi_a(t)-\phi_b(t))\delta\theta_i^a
    +\eta_i^a.
\end{eqnarray}
As Eq.~\eqref{SIeq: fluctuation dynamics} is linear, the probability distribution function $\rho_i^a(\delta\theta_i^a)$ can be computed analytically through a standard approach of mapping the Langevin equation to the Fokker-Planck equation \cite{vanKampen2003} as \cite{Lillo2000,Fa2016},
\begin{eqnarray}
    \rho_i^a(t,\delta\theta_i^a;\phi(t))
    =\frac{1}{\sqrt{\pi }w_a(t;\phi(t))}
    e^{-(\delta\theta_i^a )^2/w_a^2(t;\phi(t))}
    \nonumber\\
\end{eqnarray}
with its width $w_a$ given by,
\begin{eqnarray}
    \label{SIeq: width}
    w_a^2(t;\phi(t))
    = 2\sigma \int _0^t d\tau e^{-2\int_{\tau}^t d\tau' 
    \sum_b j_{ab}
    \cos(\phi_a(\tau')-\phi_b(\tau'))}
    \nonumber
\end{eqnarray}
when an initial condition is a perfectly magnetized state, $\delta\theta_i^a(t=0) = 0$. 
Especially in the case where $\Delta\phi_{ab}=\phi_a-\phi_b$ converges to a constant value (which occurs,  e.g., in a geometrically frustrated system and two-community perfectly non-reciprocal system), the steady-state distribution has the width \cite{vanKampen2003}
\begin{eqnarray}
    \label{SIeq: Methods width stationary}
    w_a^2(t\rightarrow\infty,\phi)= \frac{\sigma}{\sum_b j_{ab}\cos\Delta\phi_{ab}}.
\end{eqnarray}

Let us now write down the order parameter dynamics that are affected by the above fluctuations induced by noise. 
From 
\begin{eqnarray}
    \dot\psi_a = (\dot r_a + r_a i\dot\phi_a)e^{i\phi_a}
    =\frac{i}{N_a}\sum_{i=1}^{N_a}\dot\theta_i^a e^{i\theta_i^a},
\end{eqnarray}
one obtains,
\begin{eqnarray}
\label{SIeq: renormalized phase dynamics}
    \dot \phi_a
    &=&-\sum_b\frac{j_{ab}}{N_a}\sum_{i=1}^{N_a} \frac{r_b}{r_a}\sin(\theta_i^a-\phi_b)\cos(\theta_i^a-\phi_a)
    +\bar\eta_a
    \nonumber\\
    &= &  
    - \sum_b j_{ab}^\star(\phi(t))\sin(\phi_a-\phi_b)
    +\bar\eta_a
\end{eqnarray}
that is governed by the renormalized couplings,
\begin{eqnarray}
    \label{SIeq: renormalized coupling}
    &&j_{ab}^\star(\phi(t))
    = j_{ab}
    \frac{r_b(\phi(t))}{r_a(\phi(t))}
    \avg{
    \cos^2\delta\theta_i^a
    }_{\phi(t)},
\end{eqnarray}
which are, crucially, $\phi$-dependent.
Here, the effective noise for the macroscopic angle $\phi_a$ is given by $\bar\eta_a = 1/(r_a N_a)\sum_{i=1}^{N_a}\eta_a^i\cos\delta\theta_i^a\approx (1/ N_a)\sum_{i=1}^{N_a}\eta_i^a$ that
follows $\avg{\bar\eta_a}\approx 0$, $\avg{\bar\eta_a(t)\bar\eta_b(t')}\approx(\sigma/N_a)\delta_{ab}\delta(t-t')$, and  $\avg{h(\delta\theta_i^a)}_{\phi(t)}=\int d\theta_i^a \rho_i^a(t,\delta\theta_i^a;\phi(t))h(\delta\theta_i^a)$ is the noise average. 
In the second line, we have assumed that the system self-averages, i.e., $\avg{h(\delta\theta_i^a)}_{\phi(t)} = (1/N_a)\sum_{i=1}^{N_a}h(\delta\theta_i^a(t))$ and used the property $\rho_i^a(\delta\theta_i^a)=\rho_i^a(-\delta\theta_i^a)$.
As one sees by comparing with the deterministic case (Eq.~\eqref{eq: phase order dynamics deterministic} in the main text), we find that the bare couplings $j_{ab}$ has been replaced by the renormalized, $\phi$-dependent coupling  $j_{ab}^\star(\phi)$. 
For latter use,  we expand Eq.~\eqref{SIeq: renormalized phase dynamics} in terms of $\delta\theta_i^a$, giving
\begin{eqnarray}
    &&j_{ab}^\star(\phi(t))
    = j_{ab}
    \frac{\avg{\cos\delta\theta_i^b}_{\phi(t)}}
    {\avg{\cos\delta\theta_i^a}_{\phi(t)}}
    \avg{
    \cos^2\delta\theta_i^a
    }_{\phi(t)}
    \nonumber\\
    &&\simeq j_{ab}
    \frac{1-\frac{1}{2!}\avg{(\delta\theta_i^b)^2}_{\phi}
    +\frac{1}{4!}\avg{(\delta\theta_i^b)^4}_{\phi}}
    {1-\frac{1}{2!}\avg{(\delta\theta_i^a)^2}_{\phi}
    +\frac{1}{4!}\avg{(\delta\theta_i^a)^4}_{\phi}}
    \nonumber\\
    &&\times
    \Big[
    1-\avg{(\delta\theta_i^a)^2}_{\phi}
    +\frac{1}{3}
    \avg{(\delta\theta_i^a)^4}_{\phi}
    \Big]
    \nonumber\\
    &&\simeq j_{ab}
    \Big[
    1-\frac{1}{4}(w_a^2(\phi)+w_b^2(\phi))
    \nonumber\\
    &&\ \ \ \ \ \ \ \ \ \ 
    +\frac{1}{32}(5w_a^4(\phi)+2w_a^2(\phi) w_b^2(\phi)+w_b^4(\phi))
    \Big]
    \nonumber\\&&\simeq j_{ab}
    \Big[
    W(\phi)+\frac{w_a^4(\phi)}{8}
    \Big],
    \label{SIeq: renormalized coupling approx}
\end{eqnarray}
where
\begin{eqnarray}
    W(\phi)
    &=&
    1-\frac{1}{4}(w_a^2(\phi)+w_b^2(\phi))
    \nonumber\\
    &+&\frac{1}{32}(4w_a^4(\phi)+2w_a^2(\phi) w_b^2(\phi)+w_b^4(\phi)).
\end{eqnarray}
Here, we have used the relation  $r_a(\phi)=\avg{\cos\delta\theta_i^a}_{\phi}$ in the first line, expanded in terms of $\delta\theta_i^a$ in the second, and the result of the Gaussian integral
\begin{eqnarray}
    \avg{(\delta\theta_a)^2}_{\phi}
    =\frac{w_a^2(\phi)}{2},
    \qquad
    \avg{(\delta\theta_a)^4}_{\phi}
    =\frac{3w_a^4(\phi)}{4},
\end{eqnarray}
in the third. 


Below, we will show that Eq.~\eqref{SIeq: renormalized phase dynamics} generically exhibit an OBDP in both geometrically and non-reciprocally frustrated systems.

\subsubsection{Geometrically frustrated case: communities on a tetrahedron lattice}
\label{Appendix subsubsection: geometrically frustrated case}


Consider first a geometrically frustrated system that is composed of four communities, which is all-to-all antiferromagnetically coupled $j_{ab}=-j<0$ (Fig. \ref{fig: order-by-disorder}(a) in the main text ). 
We set the intra-community ferromagnetic coupling strength to be identical $j_{aa}=j_0>0$ with $a,b={\rm A,B,C,D}$ and $N_a=N$, for simplicity. 
In the absence of noise, the system is driven towards its energy minimum that are accidentally degenerate because of the geometrical frustration. 
To see this, define $\bm S_a = (S_a^x,S_a^y)=(\cos\phi_a,\sin\phi_a)$ and observe that \cite{Moessner2006,Moessner1998a,Moessner1998b} the energy $E$ can be written as,
\begin{eqnarray}
    E(\phi)=j\sum_{a,b}\bm S_a\cdot\bm S_b
    =j\Big(\sum_a \bm S_a\Big)^2+{\rm const.}
\end{eqnarray}
The ground state is given by the configuration that makes $\sum_a \bm S_a$ vanish.
As illustrated in Fig. \ref{fig: order-by-disorder}(a) in the main text, for the case considered here, 
the ground state is accidentally degenerate and is parameterized by an angle $\alpha$ and $\beta$ as
\begin{eqnarray}
\label{SIeq: ground state configuration geometrical}
    \phi_{\rm A*}=\beta,     
    \ \ 
    \phi_{\rm B*}=\pi+\beta,
    \ \ 
    \phi_{\rm C*}=\alpha+\beta,
    \ \ 
    \phi_{\rm D*}=\alpha+\pi+\beta,
    \nonumber\\
\end{eqnarray}
where the angle $\beta$ parameterizes the degeneracy trivially arising from the rotation symmetry, while $\alpha$ parameterizes the accidental degeneracy arising from geometrical frustration.
The labels of the communities can be permuted. 

Now, in the presence of noise ($\sigma>0$), the width is given by (see Eq.~\eqref{SIeq: Methods width stationary}),
\begin{eqnarray}
    w_a^2(\phi)
    &=& \frac{\sigma}{j_0 + j(\cos\pi+\cos\alpha+\cos(\alpha+\pi))}
    \nonumber\\
    &=& \frac{\sigma}{j_0 + j}
\end{eqnarray}
which is independent of the configuration $\alpha$ and is identical for all communities.
As a result, from Eq.~\eqref{SIeq: renormalized coupling approx}, one finds that $j_{ab}^\star(\phi)=-j^\star={\rm const.}<0$ on the ground state manifold, giving the macroscopic angle dynamics, ($a,b={\rm A,B,C,D}$)
\begin{eqnarray}
    \dot\phi_a = j^\star\sum_{b(\ne a)}\sin(\phi_b-\phi_a) + \bar\eta_a,
\end{eqnarray}
where $\avg{\bar\eta_a(t)}=0$ and $\avg{\bar\eta_a(t)\bar\eta_b(t')}=(\sigma/N)\delta_{ab}\delta(t-t')$.
Since this system obeys the fluctuation-dissipation theorem \cite{vanKampen2003},
the system is mapped to a problem of four spins at very low but \textit{finite} temperature $T\sim\sigma/N$. 
As we will derive below  (which is pointed out in Ref.~\cite{Moessner1998b} Sec. IV), the distribution function for realizing the angle $\alpha$ in such a system is given by Eq.~\eqref{eq: tetrahedron order by disorder} in the main text, reproduced below for convenience,
\begin{eqnarray}
    \label{SIeq: distribution geometrical}
    \rho_{ss}(\alpha)\propto \frac{1}{|\sin\alpha|},
\end{eqnarray}
in the regime $\sin^2\alpha\gg \sigma/(N|j^\star|)\rightarrow 0$.
This shows that the probability distribution is overwhelmingly concentrated to a collinear configuration $\alpha_*=0$ or $\alpha_*=\pi$,
which is nothing but an OBDP.
This is attributed to the property that, while the energy in generic configurations varies quadratically in displacement from the ground state configuration, there exists a special direction of displacement around the collinear configuration $\alpha=0,\pi$ that the energy varies \textit{quartically} \cite{Moessner1998b}, making the fluctuations in the collinear configuration large and therefore the entropy large.

The first step to derive Eq.~\eqref{SIeq: distribution geometrical} is to linearize the equation of motion around the ground state configuration $\phi_*$ (given by Eq.~\eqref{SIeq: ground state configuration geometrical}),
\begin{eqnarray}
        \delta\dot{\vec\phi}&=&\hat L(\alpha)
    \delta\vec\phi
    +\vec\eta,
\end{eqnarray}
where 
\begin{eqnarray}    
    \hat L(\alpha)=j^\star
    \begin{pmatrix}
        -1 & 1 & -\cos\alpha & \cos\alpha \\
        1 & -1 & \cos\alpha & -\cos\alpha \\
        -\cos\alpha & -\cos\alpha & -1 & 1 \\
        \cos\alpha & -\cos\alpha & 1 & -1 
    \end{pmatrix}
    \nonumber\\
\end{eqnarray}
characterizes the fluctuation dynamics,
$\delta\vec\phi=(\delta\phi_{\rm A},\delta\phi_{\rm B},\delta\phi_{\rm C},\delta\phi_{\rm D})^{\mathrm{T}}=(\phi_{\rm A},\phi_{\rm B},\phi_{\rm C},\phi_{\rm D})^{\mathrm{T}}-(\phi_{\rm A*},\phi_{\rm B*},\phi_{\rm C*},\phi_{\rm D*})^{\mathrm{T}}$ is the fluctuation, and $\vec\eta=(\bar\eta_{\rm A},\bar\eta_{\rm B},\bar\eta_{\rm C},\bar\eta_{\rm D})$ 
is the noise.

There are two zero (marginal) modes, corresponding to fluctuation within the degenerate ground state manifold.
One is the Nambu-Goldstone mode $\delta\vec\phi=(1,1,1,1)^{\mathrm{T}}$ corresponding to global rotation (therefore changing the parameter $\beta$ in Eq.~\eqref{SIeq: ground state configuration geometrical}), 
while the other $\delta\vec\phi=(1,1,-1,-1)^{\mathrm{T}}$ corresponds to changing the parameter $\alpha$.
The finite noise that continuously excites these zero modes gives rise to the diffusion of the probability distribution  of $\alpha$ and $\beta$.
However,  in the thermodynamic limit $N\rightarrow \infty$, the noise level and the diffusion constant are negligibly small.
In what follows, we will consider the timescales where this diffusion process is negligible.

On top of the two zero modes, there are two relaxational modes; one mode $(1,-1,-1,1)^{\mathrm{T}}$ with a relaxation rate $\lambda_1 = - 2j^\star (1-\cos\alpha)$
and another mode 
$(1,-1,1,-1)^{\mathrm{T}}$ with a relaxation rate $\lambda_2 = - 2j^\star (1+\cos\alpha)$.
The steady-state distribution of these fluctuation modes (denoted by $\delta\phi_1$ and $\delta\phi_2$) is then given by
\begin{eqnarray}
    \rho_{ss}(\delta\phi_1,\delta\phi_2;\alpha)
    \propto 
    e^{-4j^\star N(1-\cos\alpha)\delta\phi_1^2/\sigma}
    e^{-4j^\star N(1+\cos\alpha)\delta\phi_2^2/\sigma}.
    \nonumber\\
\end{eqnarray}
Integrating out these fluctuations, we arrive at the distribution function to realize the angle $\alpha$ as
\begin{eqnarray}
    \rho_{ss}(\alpha)&=&\int_{-\pi}^\pi d\phi_1
    \int_{-\pi}^\pi d\phi_2
    \rho_{ss}(\delta\phi_1,\delta\phi_2;\alpha)
    \nonumber\\
    &\approx&
    \int_{-\infty}^\infty d\phi_1
    \int_{-\infty}^\infty d\phi_2
    \rho_{ss}(\delta\phi_1,\delta\phi_2;\alpha)
    \nonumber\\
    &\propto&
    \int_{-\infty}^\infty d\phi_1
    e^{-4j^\star N(1-\cos\alpha)\delta\phi_1^2/\sigma}
    \nonumber\\
    &\times& 
    \int_{-\infty}^\infty d\phi_2
    e^{-4j^\star N(1+\cos\alpha)\delta\phi_2^2/\sigma}
    \nonumber\\
    &\sim &
    \frac{1}{\sqrt{1-\cos\alpha}\sqrt{1+\cos\alpha}}
    =\frac{1}{|\sin\alpha|},
\end{eqnarray}
Here, note that the approximations that are employed here are justified when $(\sigma/N)\ll|\lambda_{1,2}|$ or $\sin^2\alpha\gg \sigma/(N|j^\star|)\rightarrow 0$.
This derives Eq.~\eqref{SIeq: distribution geometrical}.

\begin{figure*}
  \centering
  \includegraphics[width=1\linewidth]{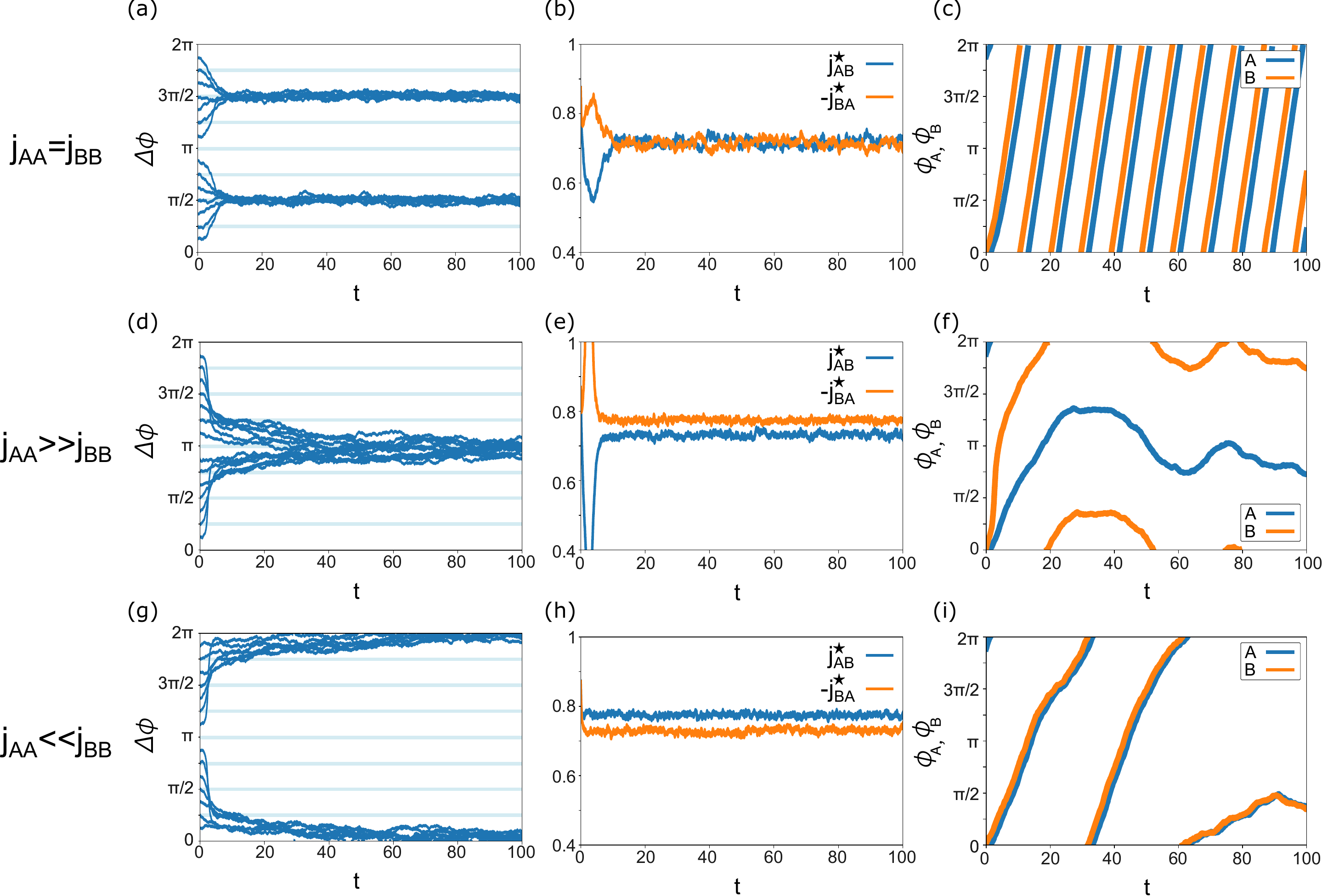}
  \caption{\textbf{Time-crystalline order-by-disorder phenomena  with different intra-community coupling strength.}
  The intra-community coupling strength is set to (a)-(c) $j_{\rm AA}=j_{\rm BB}=3$, (d)-(f) $j_{\rm AA}=5,j_{\rm BB}=2$, (g)-(i) $j_{\rm AA}=2,j_{\rm BB}=5$.
  (a),(d),(g) angle difference $\Delta\phi$ dynamics, where solid (thin) line represents the dynamics in the presence (absence) of noise. (b),(e),(h) Effective coupling $j_{ab}^\star(\phi)$. (c),(f),(i) Phase $\phi_a$ dynamics. While in (a)-(c), the chiral phase with $\Delta\phi_*=\pm\pi/2$ that satisfies $j_{\rm AB}^\star(\Delta\phi_*)=-j_{\rm BA}^\star(\Delta\phi_*)$ is ``selected'', in (d)-(f) [(g)-(i)], as the effective coupling $j_{\rm BA}^\star(\Delta\phi)[j_{\rm AB}^\star(\Delta\phi)]$ is more strongly renormalized than $j_{\rm AB}^\star(\Delta\phi)[j_{\rm BA}^\star(\Delta\phi)]$, one always finds $j_{\rm AB}^\star(\Delta\phi)<-j_{\rm BA}^\star(\Delta\phi)$  [$j_{\rm AB}^\star(\Delta\phi)>-j_{\rm BA}^\star(\Delta\phi)$ ] that stabilizes the anti-aligned [aligned] phase characterized by the phase difference $\Delta\phi_*=\pi$ [$\Delta\phi_*=0$].
  These results are all consistent with our analytical analysis (Eq.~\eqref{SIeq: phase order dynamics renormalized}). We set the noise strength $\sigma=1.5$, the number of spins $N_{\rm A}=N_{\rm B}=2000$, and the inter-community coupling strength $j_{\rm AB}=-j_{\rm BA}=1$.} 
\label{Extended figure: different jaa}
\end{figure*}

\subsubsection{Non-reciprocally frustrated case: two-community stochastic XY-model} \label{subsec:non-reciprocally frustrated}
We now turn to the non-reciprocally frustrated case. 
From Eq.~\eqref{SIeq: renormalized phase dynamics}, the dynamics of the angle difference $\Delta\phi$ that characterizes the spontaneous parity breaking is given by,
\begin{eqnarray}
    \label{SIeq: phase order dynamics renormalized}
    \Delta\dot\phi
    &=&-(j_{\rm AB}^\star(\Delta\phi)+j_{\rm BA}^\star(\Delta\phi))
    \sin\Delta\phi,
    \nonumber\\
    &\simeq &   
    -[2j_+W(\Delta\phi)+\frac{j_-}{4}(w_{\rm A}^4(\Delta\phi)
    -w_{\rm B}^4(\Delta\phi))]
    \sin\Delta\phi ,
    \nonumber\\
\end{eqnarray}
with the width 
\begin{eqnarray}
\label{SIeq: wA}
    w_{\rm A}^2(\Delta\phi)&=&\frac{\sigma}{j_{\rm AA}+j_-\cos\Delta\phi},
    \\
\label{SIeq: wB}
    w_{\rm B}^2(\Delta\phi)&=&\frac{\sigma}{j_{\rm BB}-j_-\cos\Delta\phi}.
\end{eqnarray}
Note that the second term in Eq.~\eqref{SIeq: phase order dynamics renormalized} (i.e. the entropic torque that arises only when the non-reciprocity $j_-$ and the noise $\sigma>0$ are both turned on) becomes dominant in the perfectly non-reciprocal case $j_+=0$.

We will focus here on the strong non-reciprocal case, where the reciprocal component of the inter-community coupling $|j_+|$ is smaller than the non-reciprocal part $|j_-|$, i.e. $|j_+|\ll |j_-|$. 
In particular, we will be focusing on the regime where the non-reciprocity induced entropic force (the second term of Eq.~\eqref{SIeq: phase order dynamics renormalized}) can become comparable to the reciprocal component (the first term  Eq.~\eqref{SIeq: phase order dynamics renormalized}) at small noise strength.
By expanding Eq.~\eqref{SIeq: phase order dynamics renormalized} with respect to $j_+$, the $\Delta\phi$-dynamics reads
\begin{eqnarray}
\label{SIeq: Delta phi jAA=jBB}
    \Delta\dot\phi
    &\simeq &
    \Big[
    -2j_+
    +\frac{j_0 j_-^2\sigma^2}{2}
    \frac{ \cos\Delta\phi }{(j_0^2 - j_-^2 \cos^2\Delta\phi)^2}
    \Big]
    \sin\Delta\phi,
    \nonumber\\
\end{eqnarray}
for the identical intra-community coupling case $j_{\rm AA}=j_{\rm BB}=j_0$,
deriving Eq.~\eqref{eq:  phase order dynamics renormalized small j+} (and Eq.~\eqref{eq:  Delta phi jAA=jBB} for the case of $j_+=0$) in the main text.
As  thoroughly discussed in the main text Sec.~\ref{sec:order-by-disorder}, the non-reciprocity induced entropic torque triggers a non-reciprocal phase transition~\cite{Fruchart2021} to a chiral phase $\Delta\phi_*\ne 0,\pi$ when the noise strength $\sigma$ exceeds a critical value.




It is worth noting that the non-reciprocal frustration-induced torque does not always prefer the chiral phase (that spontaneously breaks parity).
To make the discussion simple, let us consider below the perfectly non-reciprocal case $j_+=0$.
When the intra-community coupling of community A(B) is sufficiently large compared to B(A), i.e., $j_{\rm AA}\gg j_{\rm BB}(j_{\rm AA}\ll j_{\rm BB})$, the community A(B) becomes stiff such that the fluctuations of the community A(B) get strongly suppressed to give $w_{\rm A}^2(\Delta\phi)< w_{\rm B}^2(\Delta\phi)(w_{\rm A}^2(\Delta\phi)> w_{\rm B}^2(\Delta\phi))$ for \textit{arbitrary} $\Delta\phi$ (See Eqs.~\eqref{SIeq: wA} and \eqref{SIeq: wB}.).
As a result, no configuration can satisfy the condition that 
$j_{\rm AB}^\star(\Delta\phi)+j_{\rm BA}^\star(\Delta\phi)
\simeq    
-\frac{j_-}{4}(w_{\rm A}^4(\Delta\phi)
-w_{\rm B}^4(\Delta\phi))$ must vanish in the chiral phase (see Eq.~\eqref{eq:  chiral phase condition} in the main text).
In this case, the effective coupling always satisfies $j_{\rm AB}^\star(\Delta\phi)<|j_{\rm BA}^\star(\Delta\phi)|(j_{\rm AB}^\star(\Delta\phi)>|j_{\rm BA}^\star(\Delta\phi)|)$ for arbitrary $\Delta\phi$ when $j_->0$. 
This leads the system to select $\Delta\phi_*=\pi(0)$ for $j_->0$, corresponding to a static phase $\dot{\Phi}_*(t)= 0$, even in the absence of bare reciprocal coupling $j_+=0$.
All these features are demonstrated numerically in Figs.~\ref{Extended figure: different jaa}(b),(c). 

Similar features are obtained when the noise strength is different between different communities, i.e. when the noise is characterized by $\avg{\eta_a(t)}=0,\avg{\eta_a(t)\eta_b(t')}=\sigma_{ab}\delta_{ab}\delta(t-t')$ with $\sigma_{\rm A}\ne \sigma_{\rm B}$.
In this case, the width of fluctuation of each community is given by 
\begin{eqnarray}
\label{SIeq: wA sigmaA}
    w_{\rm A}^2(\Delta\phi)&=&\frac{\sigma_{\rm A}}{j_{\rm AA}+j_-\cos\Delta\phi},
    \\
\label{SIeq: wB sigmaB}
    w_{\rm B}^2(\Delta\phi)&=&\frac{\sigma_{\rm B}}{j_{\rm BB}-j_-\cos\Delta\phi},
\end{eqnarray}
leading to a similar situation to the above when the noise strength is sufficiently different between the two communities.
For example, when $\sigma_{\rm A}\gg\sigma_{\rm B}$ ($\sigma_{\rm A}\ll\sigma_{\rm B}$) and $j_{\rm AA}=j_{\rm BB}=j_0$,
$w_{\rm A}^2(\Delta\phi)>w_{\rm B}^2(\Delta\phi)$ 
($w_{\rm A}^2(\Delta\phi)<w_{\rm B}^2(\Delta\phi)$)
would always be satisfied for arbitrary $\Delta\phi$, leading to a static phase with (anti-)aligned configuration $\Delta\phi_*=0$ ($\Delta\phi_*=\pi$).



\subsection{Spatially extended model}


We consider here the spatially extended model governed by Eqs.~\eqref{eq: Langevin d-dim A} and \eqref{eq: Langevin d-dim B} (and illustrated in Fig.~\ref{fig: model spatial}) in the main text.
In the presence of noise $\sigma>0$,
the angle fluctuates around its mean value as $\delta\theta_{\bm x}^a = \theta_{\bm x}^a - \phi_a$,
where the macroscopic angle $\phi_a$ is defined from the order parameter $\psi_a=r_a e^{i\phi_a}
=(1/N)\sum_{\bm x} e^{i\theta_{\bm x}^a}$ dynamics for the sublattice $a={\rm A,B}$. 
In the Fourier space, fluctuations $\delta\theta_{a,\bm k}$ obey
\begin{eqnarray}
    \partial_t \begin{pmatrix}
        \delta\theta_{{\rm A},\bm k} \\
        \delta\theta_{{\rm B},\bm k} 
    \end{pmatrix}
    =\hat L_{\bm k}(\Delta\phi)
    \begin{pmatrix}
        \delta\theta_{{\rm A},\bm k} \\
        \delta\theta_{{\rm B},\bm k} 
    \end{pmatrix}
    +\begin{pmatrix}
        \eta_{{\rm A},\bm k} \\
        \eta_{{\rm B},\bm k} 
    \end{pmatrix},
\end{eqnarray}
where noise is characterized by
$\avg{\eta_{a,\bm k}(t)}=0$, $\avg{\eta_{a,\bm k}(t)\eta_{b,\bm k'}(t')}=\sigma(2\pi)^{d+1} \delta_{a,b}\delta(t-t')\delta^{d}(\bm k+\bm k')$ and
the dynamical matrix is given in the case of $d=2$ as
\begin{widetext}
    \begin{eqnarray}
    \hat L_{\bm k}(\Delta\phi)
    &=&
    \begin{pmatrix}
        j_{\rm AA}(\cos k_x \cos k_y -1)-j_{\rm AB}\cos\Delta\phi &
        (j_{\rm AB}/2)\cos\Delta\phi
         (\cos k_x + \cos k_y) \\
        (j_{\rm BA}/2)\cos\Delta\phi
         (\cos k_x + \cos k_y) &
         j_{\rm BB}(\cos k_x \cos k_y -1)
         -j_{\rm BA}\cos\Delta\phi
        \\
    \end{pmatrix}
    \nonumber\\
    &\simeq &
    \begin{pmatrix}
        -\frac{j_{\rm AA}}{2}\bm k^2
        -j_{\rm AB}\cos\Delta\phi &
        j_{\rm AB}\cos\Delta\phi
         (1-\frac{\bm k^2}{2}) \\
        j_{\rm BA}\cos\Delta\phi
         (1-\frac{\bm k^2}{2}) &
         -\frac{j_{\rm BB}}{2}\bm k^2
         -j_{\rm BA}\cos\Delta\phi
        \\
    \end{pmatrix}.
\end{eqnarray}
The extension to higher spatial dimension $d>2$ is straightforward, and the final expression is applicable for arbitrary spatial dimensions.
As in the previous sections, through a standard approach of mapping the Langevin equation to the Fokker-Planck equation \cite{vanKampen2003}, the distribution function $\rho_{ss}(\{\delta\theta_{{\rm A},\bm k},\delta\theta_{{\rm B},\bm k}\};\Delta\phi)
=\prod_{\bm k}\rho_{ss,\bm k}(\delta\theta_{{\rm A},\bm k},\delta\theta_{{\rm B},\bm k};\Delta\phi)$ can be obtained as a product of Gaussian distribution, 
\begin{eqnarray}
    \rho_{ss,\bm k}(\delta\theta_{{\rm A},\bm k},\delta\theta_{{\rm B},\bm k};\Delta\phi)
    = \frac{1}{\sqrt{(2\pi){\rm det}\hat\Xi_{\bm k}}}
    \exp[-\frac{1}{2}
    (\hat\Xi_{\bm k}^{-1}(\Delta\phi))_{ab}
    \delta\theta_{\bm k,a}\delta\theta_{-\bm k,b}]
\end{eqnarray}
where the correlation matrix is given by (where $j_\pm = (j_{\rm AB}\pm j_{\rm BA})/2$),
\begin{eqnarray}
    &&\hat\Xi_{\bm k}(\Delta\phi)
    \simeq \frac{\sigma}{\bm k^2 (2  j_+\cos\Delta\phi + j_0\bm k^2)(4j_0 j_+\cos\Delta\phi 
    -4\cos^2\Delta\phi(j_-^2-j_+^2)+j_0^2\bm k^2)}
    \nonumber\\
    &&\times
    \begin{pmatrix}
       j_0^2\bm k^4 -2\cos\Delta\phi j_0(j_--2j_+)\bm k^2 +  4\cos^2\Delta\phi (j_-^2+j_+^2) 
       & 
       2\cos\Delta\phi(2\cos\Delta\phi(j_-^2+j_+^2)+j_-j_+\bm k^2)
       \\
       2\cos\Delta\phi(2\cos\Delta\phi(j_-^2+j_+^2)+j_-j_+\bm k^2)
       & j_0^2\bm k^4 +2\cos\Delta\phi j_0(j_--2j_+)\bm k^2 +  4\cos^2\Delta\phi (j_-^2+j_+^2 ) 
    \end{pmatrix}.
    \nonumber\\
\end{eqnarray}
We have restricted ourselves to the case with $j_{\rm AA}=j_{\rm BB}=j_0$ for simplicity.
This gives the variance of fluctuations as
\begin{eqnarray}
\label{SIeq: variance spatial}
    \avg{\delta\theta_{\bm x}^a\delta\theta_{\bm x}^b}_{\Delta\phi}&=&\sum_{\bm k}\int_{-\infty}^\infty d\delta\theta_{{\rm A},\bm k}
    \int_{-\infty}^\infty d\delta\theta_{{\rm B},\bm k}
    \rho_{ss,\bm k}(\delta\theta_{{\rm A},\bm k},\delta\theta_{{\rm B},\bm k};\Delta\phi)
    \delta\theta_{a,\bm k}
    \delta\theta_{b,-\bm k}
    =\frac{1}{4}\sum_{\bm k}\Xi_{\bm k,ab}(\Delta\phi).
\end{eqnarray}
We briefly note that, while generically the correlation matrix is inversely proportional to $\bm k^2$ ($\hat\Xi_{\bm k}\propto \bm k^{-2}$), they behave as $\hat\Xi_{\bm k}\propto \bm k^{-4}$ in the case of perfect non-reciprocity $j_+=0$.
This feature, which implies the diverging fluctuations $\avg{(\delta\theta_{\bm x}^a)^2}\sim \int dk k^{d-1}|\delta\theta_{\bm k, a}|^2\sim\int dk k^{d-1}k^{-4}\rightarrow\infty$ at $d<4$ (implying the destruction of long-range order at $d<4$) in the vicinity of a critical point, is a characteristic of critical exceptional point which is the salient feature of  non-reciprocal phase transitions \cite{Hanai2020, Zelle2023}.

Equipped with the distribution of fluctuations, we now consider their impact on the macroscopic phase $\phi_a$ dynamics. 
The dynamics of the order parameter reads
\begin{eqnarray}
    \dot\psi_{a} 
    = (\dot r_{a}+i r_{a}\dot\phi_{a})
    e^{i\phi_{a}}
    =i\sum_{\bm x}\dot\theta_{\bm x}^{a}
    e^{i\theta_{\bm x}^{a}}.
\end{eqnarray}
Plugging in the governing equation \eqref{eq: Langevin d-dim A} and \eqref{eq: Langevin d-dim B} in the main text, we arrive at the same form as Eq.~\eqref{eq: order parameter renormalized} in the main text, reproduced below for convenience:
\begin{eqnarray}
    \label{SIeq: order parameter renormalized}
    \dot\phi_a(t) 
    = -\sum_b j_{ab}^\star(\phi(t))
    \sin(\phi_a(t)-\phi_b(t))
    +\bar\eta_a(t).
\end{eqnarray}
Here, the renormalized coupling given by
\begin{eqnarray}
    \label{SIeq: jABstar short-range}
    &&j_{\rm AB}^\star(\Delta\phi)
    =\frac{j_{\rm AB}}{r_{\rm A}(\Delta\phi)}
    \avg{
    \cos(\delta\theta_{\bm x+\hat b}^{\rm B}
    -\delta\theta_{\bm x}^{\rm A})
    \cos\delta\theta_{\bm x}^{\rm A}
    }_{\Delta\phi},
    \\
    \label{SIeq: jBAstar short-range}
    &&j_{\rm BA}^\star(\Delta\phi)
    =\frac{j_{\rm BA}}{r_{\rm B}(\Delta\phi)}
    \avg{
    \cos(\delta\theta_{\bm x+\hat b}^{\rm A}
    -\delta\theta_{\bm x}^{\rm B})
    \cos\delta\theta_{\bm x}^{\rm B}
    }_{\Delta\phi},
\end{eqnarray}
where
$\avg{\cdots}_{\Delta\phi}=\int \prod d\delta\theta_{\bm x}^{\rm A}\delta\theta_{\bm x}^{\rm B}
\rho(\{\delta\theta_{\bm x}^{\rm A},\delta\theta_{\bm x}^{\rm B}\};\Delta\phi)(\cdots)$ is the noise average with the configuration $\Delta\phi$ 
(where $\rho(\{\delta\theta_{\bm x}^{\rm A},\delta\theta_{\bm x}^{\rm B}\};\Delta\phi)$ is the distribution to realize $(\delta\theta_{\bm x}^{\rm A},\delta\theta_{\bm x}^{\rm B})$ with $\Delta\phi$) and $r_a(\Delta\phi)=\avg{\cos\delta\theta_{\bm x}^a}_{\Delta\phi}$. $\bar\eta_a$ is a macroscopic noise that has the noise strength of $\sigma/N$.
We have assumed that the system self-averages.

We proceed by expanding Eqs.~\eqref{SIeq: jABstar short-range} and \eqref{SIeq: jBAstar short-range} in terms of fluctuations as
\begin{eqnarray}
    j_{\rm AB}^\star(\Delta\phi)
    &\simeq&
    j_{\rm AB}\big[
    W(\Delta\phi)+\frac{1}{4}
    \avg{(\delta\theta_{\bm x}^{\rm A})^2}_{\Delta\phi}^2
    \big],\\
    j_{\rm BA}^\star(\Delta\phi)
    &\simeq &
    j_{\rm BA}
    \big[
    W(\Delta\phi)+\frac{1}{4}
    \avg{(\delta\theta_{\bm x}^{\rm B})^2}_{\Delta\phi}^2
    \big],
\end{eqnarray}
where
\begin{eqnarray}
    &&W(\Delta\phi)
    =1 - \frac{1}{2}
    \Big[\avg{(\delta\theta_{\bm x}^{\rm A})^2}_{\Delta\phi}
    +\avg{(\delta\theta_{\bm x}^{\rm B})^2}_{\Delta\phi}
    -2\avg{\delta\theta_{\bm x}^{\rm A}\delta\theta_{\bm x+\hat b}^{\rm B}}_{\Delta\phi}
    \Big]
    \nonumber\\
    &&+\frac{1}{24}
    \Big[\avg{(\delta\theta_{\bm x}^{\rm A})^4}_{\Delta\phi}
    +\avg{(\delta\theta_{\bm x}^{\rm B})^4}_{\Delta\phi}
    +12\avg{(\delta\theta_{\bm x}^{\rm A})^2
    (\delta\theta_{\bm x}^{\rm B})^2}_{\Delta\phi}
    -4\avg{(\delta\theta_{\bm x}^{\rm A})^3\delta\theta_{\bm x+\hat b}^{\rm B}}_{\Delta\phi}
    -4\avg{\delta\theta_{\bm x}^{\rm A}(\delta\theta_{\bm x+\hat b}^{\rm B})^3}_{\Delta\phi}
    \Big].
\end{eqnarray}  
The dynamics of $\Delta\phi$, which characterizes the parity breaking order of the chiral phase  \cite{Fruchart2021}, then follows,
\begin{eqnarray}
    &&\Delta\dot\phi
    =-(j_{\rm AB}^\star(\Delta\phi)
    +j_{\rm BA}^\star(\Delta\phi))\sin\Delta\phi
    \nonumber\\
    &&\simeq 
    -[2j_+
    W(\Delta\phi)
    +\frac{j_-}{2}
    (
    \avg{(\delta\theta_{\bm x}^{\rm A})^2}_{\Delta\phi}^2
    -
    \avg{(\delta\theta_{\bm x}^{\rm B})^2}_{\Delta\phi}^2)]\sin\Delta\phi.
\end{eqnarray}
Further assuming $j_+\ll j_-,j_0$, using Eq.~\eqref{SIeq: variance spatial}, we find
\begin{eqnarray}
\label{SIeq: Delta phi dynamics spatial}
 \Delta\dot\phi&\simeq&\bigg[-2j_+
 + \frac{ j_-^2\sigma^2
 \cos\Delta\phi}{4j_0}
 \sum_{\bm k,\bm {k'}}
 \frac{j_0^2{\bm k'}^2
 -2 j_-^2\cos^2\Delta\phi}
 {(4j_-^2\cos^2\Delta\phi-j_0^2\bm k^2)(4j_-^2\cos^2\Delta\phi-j_0^2{\bm k'}^2)\bm k^2 {\bm k'}^2}
 \bigg] \sin\Delta\phi,
\end{eqnarray}
which derives Eq.~\eqref{eq: Delta phi dynamics spatial} in the main text.
\end{widetext}

\bibliography{main}

\end{document}